\documentclass[sigconf, screen]{acmart}
\renewcommand\footnotetextcopyrightpermission[1]{}
\PassOptionsToPackage{dvipsnames}{xcolor}
\PassOptionsToPackage{colorlinks=true,linkcolor=RubineRed,breaklinks=True,citecolor=RubineRed}{xcolor}

\usepackage{enumitem}
\setlist[itemize]{noitemsep, topsep=0pt}

\usepackage[most]{tcolorbox}
\usepackage{adjustbox}
\usepackage[linesnumbered,algo2e,ruled]{algorithm2e}
\usepackage{multirow}
\usepackage{booktabs,subcaption,dcolumn}
\usepackage{tikz}
\usepackage{enumitem}
\usepackage{tabularx}
\usepackage{MnSymbol}
\usepackage{longtable}
\usepackage{microtype}

\usepackage{pifont}
\usepackage{svg}
\usepackage{soul}
\usepackage{amsmath}
\usepackage{dirtytalk}
\usepackage{subcaption}
\usepackage{graphicx}
\usepackage{setspace}
\usepackage[font={stretch=0.3, small}]{caption}
\usepackage{booktabs} 
\usepackage[titletoc,toc,title]{appendix}
\usepackage{svg}
\usepackage{amsfonts}
\usepackage{xurl} 

\usepackage{diagbox}
\usepackage{indentfirst}

\usepackage{amsmath}
\usepackage{amsfonts}

\usepackage{colortbl}

\copyrightyear{2025}
\acmYear{2025}
\setcopyright{acmlicensed}\acmConference[ASIA CCS '25]{ACM Asia Conference
on Computer and Communications Security}{August 25--29, 2025}{Hanoi,
Vietnam}
\acmBooktitle{ACM Asia Conference on Computer and Communications
Security (ASIA CCS '25), August 25--29, 2025, Hanoi, Vietnam}
\acmDOI{10.1145/3708821.3736195}
\acmISBN{979-8-4007-1410-8/2025/08}

\AtBeginDocument{%
  \providecommand\BibTeX{{%
    \normalfont B\kern-0.5em{\scshape i\kern-0.25em b}\kern-0.8em\TeX}}}

\newcommand{\cmark}{{\textcolor{ACMGreen}{\ding{51}}\xspace}} 
\newcommand{\xmark}{{\textcolor{ACMRed}{\ding{55}}\xspace}} 

\newcolumntype{?}{!{\vrule width 1.5pt}}

\newcommand{\textbox}[1]{
    \noindent\fbox{%
        \parbox{0.97\columnwidth}{%
            {#1}
        }%
    }
}

\newtcolorbox{cooltextbox}[1][]{%
    colback=black!5,
    colframe=black!5,
    notitle,
    sharp corners,
    borderline west={0pt}{0pt}{red!80!black},
    enhanced,
    breakable,
    left=0pt,
    right=0pt,
    top=0pt,
    bottom=0pt
    }

\newcommand\smamath[1]{{\small $#1$}}

\newcommand\scmath[1]{{\scriptsize $#1$}}

\newcommand{\cs}[1][\small]{{#1$\mathbb{C}_s$}}
\newcommand{\cm}[1][\small]{{#1$\mathbb{C}_m$}}
\newcommand{\ch}[1][\small]{{#1$\mathbb{C}_h$}}

\newcommand{\eb}[1][\small]{{#1$\mathbb{E}_B$}}
\newcommand{\eq}[1][\small]{{#1$\mathbb{E}_Q$}}
\newcommand{\el}[1][\small]{{#1$\mathbb{E}_L$}}

\newcommand{\prompt}[2][\small]{{\fontfamily{qcr}\selectfont {#1 #2}}}

\begin{document}

\title{The Impact of Emerging Phishing Threats: Assessing Quishing and LLM-generated Phishing Emails against Organizations}

\author{Marie Weinz}
\email{marie.weinz@uni.li}
\orcid{0009-0003-2112-5697}
\affiliation{%
  \institution{University of Liechtenstein}
  \city{Vaduz}
  \country{Liechtenstein}
}

\author{Nicola Zannone}
\email{n.zannone@tue.nl}
\orcid{0000-0002-9081-5996}
\affiliation{%
  \institution{Eindhoven University of Technology}
  \city{Eindhoven}
  \country{Netherlands}
}

\author{Luca Allodi}
\email{l.allodi@tue.nl}
\orcid{0000-0003-1600-0868}
\affiliation{%
  \institution{Eindhoven University of Technology}
  \city{Eindhoven}
  \country{Netherlands}
}

\author{Giovanni Apruzzese}
\email{giovanni.apruzzese@uni.li}
\orcid{0000-0002-6890-9611}
\affiliation{%
  \institution{University of Liechtenstein}
  \city{Vaduz}
  \country{Liechtenstein}
}

\begin{abstract}
Modern organizations are persistently targeted by phishing emails. Despite advances in detection systems and widespread employee training, attackers continue to innovate, posing ongoing threats. Two emerging vectors stand out in the current landscape: QR-code baits and LLM-enabled pretexting. 
Yet, little is known about the effectiveness of current defenses against these attacks, particularly when it comes to real-world impact on employees. This gap leaves uncertainty around to what extent related countermeasures are justified or needed. Our work addresses this issue.

We conduct three phishing simulations across organizations of varying sizes---from small-medium businesses to a multinational enterprise.
In total, we send over 71k emails targeting employees, including: a ``traditional'' phishing email with a click-through button; a nearly-identical ``quishing'' email with a QR code instead; and a phishing email written with the assistance of an LLM and open-source intelligence. 
Our results show that quishing emails have the same effectiveness as traditional phishing emails at luring users to the landing webpage---which is worrying, given that quishing emails are much harder to identify even by operational detectors. 
We also find that LLMs can be very good ``social engineers'': in one company, over 30\% of the emails opened led to visiting the landing webpage---a rate exceeding some prior benchmarks. Finally, we complement our study by conducting a survey across the organizations' employees, measuring their ``perceived'' phishing awareness. Our findings suggest a correlation between higher self-reported awareness and organizational resilience to phishing attempts.
\end{abstract}

\ccsdesc[500]{Security and privacy~Phishing}

\keywords{phishing, quishing, chatgpt, education, user study, awareness, email}

\settopmatter{printfolios=true}

\maketitle

\section{Introduction}
\label{sec:introduction}
\noindent
Phishing is endemic in the threat landscape of modern organizations. According to Proofpoint's 2024 State of the Phish report~\cite{proofpoint2024phish}, among the top-5 most prevalent attacks suffered by organizations, four revolved around phishing. Worryingly, over 66 million business-email-compromise attacks have been detected \textit{every month} in 2023, a finding that is attributed to recent advances in generative artificial intelligence (AI). The report also stated that even though nearly every company implements some form of phishing education, only 23\% educate on generative AI safety---and 
only half (53\%) provide training for everyone, leading to ``gaps'' exploitable by phishers.

Phishing remains a persistent and evolving threat, despite having been studied for decades~\cite{baki2023sixteen,dhamija2005battle}.
In response, the research community and industry have developed a variety of countermeasures, including phishing email detectors~\cite{lee2021dfence,salloum2021phishing,valecha2021phishing}, as well as simulation, training, and educational campaigns~\cite{lain2024content,schiller2024employees,ho2025understanding,lain2022phishing}.
Yet, attackers continue to bypass these defenses and achieve their objectives. This ongoing challenge stems from the absence of a universal solution: phishing exploits both the technical limitations of automated systems and the cognitive biases of human users.
Recent developments further complicate the landscape. Adversaries are increasingly using large language models (LLMs) to generate convincing phishing content~\cite{violino2023chatgpt,slashnext2023phishing,egress2024email,proofpoint2024phish}, and are leveraging QR codes as a novel delivery mechanism for phishing attacks~\cite{talos2024how,barracuda2024report,malwarebytes2025qr}.
Addressing this threat requires continuous monitoring of emerging techniques and the development of adaptive, targeted defenses.

Unfortunately, development and implementation of such countermeasures are progressing at a slow pace. Take ``quishing'' emails, for instance: despite some works proposing ways to detect malicious QR-codes in emails (e.g.,~\cite{vaithilingam2024enhancing, yong2019survey}), quishing emails still evade most ``automated'' filters---and attackers are aware of this. Recent reports by Cisco Talos~\cite{talos2024qr} state that over 60\% of the emails containing a QR code are not benign---a trend confirmed also by other recent reports~\cite{barracuda2024report}. At the same time, despite the increasing reliance on LLMs by phishers~\cite{egress2024email,violino2023chatgpt,slashnext2023phishing}, there is still a lack of education on how to spot (deceitful) LLM-written content~\cite{proofpoint2024phish}.
\looseness=-1

We argue that such deficiencies are due to an overall poor understanding of such emerging phishing threats (discussed in  §\ref{sec:background}). 
However, such technologies are now becoming an important asset for professional phishers~\cite{proofpoint2024phish}. Therefore, it is necessary to analyze the effectiveness of these emerging phishing threats against employees, and how this might be expected to vary (or not) across companies. This paper seeks to fulfill this gap.
\looseness=-1

To achieve our goal, we first find an agreement with three distinct companies: doing so enables us to gauge the extent to which our findings hold across different companies. Then, we carry out phishing simulations focusing on investigating:
{\small \textit{(i)}}~the effectiveness of QR-code phishing emails compared to traditional click-through phishing emails; and
{\small \textit{(ii)}}~the effectiveness of LLM-based phishing emails that have been fed with open-source intelligence (OSINT) information. Finally, we carry out a complementary investigation on the relationship between \textit{perceived phishing awareness} and \textit{actual phishing susceptibility} across our considered companies' employees.
Connecting all three aspects (perceived phishing awareness, actual phishing susceptibility, diverse companies) is of great value for research, since it would enrich the findings of all prior work that investigated only one or two of these aspects in an organizational context (e.g.,~\cite{lain2022phishing, reinheimer2020investigation,gordon2019assessment,yang2022predicting}), as well as provide practical suggestions for real organizations (e.g., if there is a correlation between perceived phishing awareness and actual phishing susceptibility, then one can be a predictor of the other) or providers of security services (e.g., prioritizing the development of automated countermeasures). 
\looseness=-1

\textsc{\textbf{Contributions and Findings.}} We provide factual data on how a diverse set of real-world companies deal with, and are affected by, the multi-faceted threat of phishing emails. 
We carry out a \textit{large-scale and fine-grained assessment} of employees' susceptibility to three types of targeted-phishing emails reflecting emerging trends. Our sample (\smamath{71\,309} total emails sent) refers to companies of \textit{different business size} (respectively: \smamath{<}250, \smamath{\approx}1\,500, \smamath{>}30\,000 employees). Our setup enables comparison of the results across companies.

\begin{itemize}[leftmargin=*]
    \item We scrutinize whether \textbf{QR-based phishing emails} are more (or less) effective than traditional button-based phishing emails at luring users to a malicious webpage. We find  \textit{no statistically significant} (\smamath{p}=\smamath{.552}) difference, and a TOST test of equivalence confirms that the two groups are indistinguishable at a tolerance level of less than 1\%. 
    This apparently counterintuitive result (scanning a QR code in an email is not as straightforward as clicking on a button) has unfortunately several concerning security implications (described in §\ref{sssec:quishing}, and verified by an experiment).

    \item We study the effectiveness of \textbf{combining a LLM with OSINT} to craft targeted phishing emails. 
    Our assessment reveals that even employees with prior phishing training can be deceived using freely available AI tools and public information. For instance, for the second company, \smamath{\approx}\smamath{10\%} of the recipients (n=\smamath{589}) submitted their credentials, and \smamath{\approx}\smamath{21\%} visited the webpage. 
        
    \item Through an informed survey with a subset of our companies' employees (n=\smamath{131}), we measure their degree of \textbf{perceived phishing awareness} (PPA). We then cross-analyse the PPA of each company with the overall effectiveness of our phishing simulations. 
    We find that the PPA can be a statistically significant (\smamath{p}=\smamath{.044}) predictor of the effectiveness of a phishing campaign. 
\end{itemize}
Finally, we provide all our fine-grained results in Appendix~\ref{app:questionnaires}. Such details are not only important for transparency, but are also useful for benchmarking and comparative purposes.
\section{Background and Motivation}
\label{sec:background}
\noindent
Numerous reports from cybersecurity agencies underscore the impact of phishing, and particularly phishing emails,\footnote{We focus on phishing (and not spam~\cite{janez2023review}) \textit{emails}: other forms of phishing (e.g., websites~\cite{oest2020phishtime}, SMS~\cite{jakobsson2018two,nahapetyan2024sms}, or via vocal telephony~\cite{yeboah2014phishing}) are outside our scope.} on modern organizations~\cite{proofpoint2024phish,egress2024email,apwg2024report,barracuda2024report}. 
Below, we summarize the risks posed by two emerging phishing-email threats in corporate contexts~(§\ref{ssec:trends}), and then highlight the research gap that we aim to fill~(§\ref{ssec:gap}).

\subsection{Emerging Trends in Phishing Emails}
\label{ssec:trends}
\noindent
We motivate the problem tackled in our paper by outlining the properties of QR-code phishing~(§\ref{sssec:quishing}) and LLM-based phishing~(§\ref{sssec:llm}), explaining why they are particularly problematic today.

\subsubsection{\textbf{Quishing: Social Engineering via QR codes}\\}
\label{sssec:quishing}

\noindent
QR-code phishing, also referred to as ``Quishing'', is a form of social engineering attack which attempts to deceive individuals into scanning QR codes that point to a malicious website~\cite{amoah2022qr,sharevski2024exploring}. In some cases, such malicious QR codes are spread in the real world (e.g., physically glued to objects~\cite{talos2024qr}); however, the majority of quishing incidents originate from emails~\cite{talos2024how}, which can have a malicious QR code included either as an image attachment, or embedded in the email's body. To deceive their targets, quishing emails leverage the same methods as regular phishing emails, e.g., mimicking reputable sources and emphasizing urgency \cite{hoxhunt2024quishing, krombholz2014qr}. However, two peculiarities make quishing emails particularly subtle.
\begin{itemize}[leftmargin=*]
    \item \textit{Quishing emails are not detected by spam/phishing filters.} Many providers of email services now integrate automated blocking mechanisms that prevent delivery of emails containing malicious URLs, thereby defusing most phishing attacks. However, if the malicious URL is concealed by a QR code, then such filters would not work~\cite{talos2024how}. We have  verified this property with an original experiment (discussed in Appendix~\ref{app:considerations}). Further, a user cannot evaluate the safety or trustworthiness of a QR-code in the same way they may have been trained for phishing links.
    
    \item \textit{Quishing emails bypass organization-wide security barriers.} Most workplaces adopt security mechanisms such as firewalls, VPN or private DNS. Therefore, even under the assumption that an employee receives an email with a phishing URL (potentially concealed in a click-through button~\cite{lin2019susceptibility}), clicking the URL may lead to a response of the security system---preventing the user from reaching the malicious website. However, QR codes are typically scanned with a different device, such as smartphones; such devices may not be connected to organization's network, but to that of, e.g., the telecommunication provider of the user---which operates outside the organization's security control. Hence, a user receiving a quishing email would scan the QR code and visit the linked website with an ``unprotected'' device, increasing the likelihood of falling victim to the phishing attempt~\cite{talos2024how}. 
\end{itemize}
In short, quishing emails sidestep most filters and security controls. 
We highlight two recent works that have addressed the problem of quishing (but differently from us).
Sharevski et al.~\cite{sharevski2022phishing} conducted a user study with 173 Amazon Mechanical Turk workers (i.e., humans) exposed to a fictitious malicious QR code. The goal was to determine whether users would scan the code and visit the associated URL or refrain. Only 14.5\% chose not to scan, indicating high susceptibility to quishing in this population. However, the study did not consider organizational settings or compare the effectiveness of QR-based phishing emails to traditional link-based ones, both central to our investigation.
Roy et al.~\cite{roy2024chatbots} examine the use of large language models (LLMs) to generate phishing emails, some featuring malicious QR codes. While they show that commercial LLMs (e.g., ChatGPT, Claude, Bard) can easily produce convincing quishing content, they do not assess the real-world effectiveness of these emails on human targets, especially in organizational contexts.

\subsubsection{\textbf{LLM-generated Phishing Emails}\\}
\label{sssec:llm}

Advancements in Artificial Intelligence (AI), particularly the emergence of publicly accessible large language models (LLMs), represent a double-edged sword~\cite{schroeer2025sok}. 
On one hand, LLMs enhance productivity by supporting tasks such as text generation and analysis~\cite{cambon2023early,weber2024significant,filippo2024future}. On the other hand, these same tools can be exploited by malicious actors to streamline and scale cyberattacks, lowering the barrier to crafting sophisticated offensive content.

Importantly, LLM-based tools not only are effective at {\small \textit{(i)}}~imitating human writing to create persuasive texts~\cite{frank2024representative}, but they can also {\small \textit{(ii)}}~facilitate the summarization of unstructured information~\cite{zhang2024benchmarking}, such as that acquired via open-source intelligence (OSINT). Moreover, {\small \textit{(iii)}}~LLM can now be used by anyone (essentially) for free~\cite{schroeer2025sok}. The combination of these three factors makes LLM very attractive for phishers. It is hence not surprising that numerous company executives and technical reports from renowned cyber-security companies affirm that there is an increasing usage of LLM (or ``generative AI'') to convey phishing attacks~\cite{egress2024email,violino2023chatgpt,stacey2025ai,proofpoint2024phish,slashnext2023phishing}.

Prior works have explored the potential of LLM in the phishing-email context. Most research papers (e.g.,~\cite{roy2024chatbots,langford2023phishing}, as well as various preprints~\cite{hazell2023large,hazell2023spear,karanjai2022targeted}) depict ways in which LLM can be used to craft phishing-related content. However, and to our knowledge, there are only three (unpublished, at the time of writing) works that analysed the effectiveness of LLM-based phishing emails against humans: a case report by IBM's X-Force~\cite{ibm2023llm} across 800 employees of a healthcare company evaluated the effectiveness of emails generated by feeding OSINT-acquired information (from online social networks) to an LLM (ChatGPT); Bethany et al.~\cite{bethany2024large} carried out a simulation on 9000 members of a university by asking ChatGPT to write phishing emails in a style that imitated that of some official webpages of the university; Heiding et al.~\cite{heiding2024evaluating} consider various LLMs to craft spear-phishing emails targeting one among 101 volunteers in a user study. Hence, despite the great interest in LLM in the phishing-email context, there is still a lack of understanding of how effective these tools can be at deceiving humans. (Note: using LLMs as phishing-email \textit{detectors}~\cite{heiding2024devising} is orthogonal to our work.)

\subsection{Research and Knowledge Gap}
\label{ssec:gap}

\noindent
As acknowledged in Proofpoint's latest report, QR-code and LLM-based phishing emails are becoming trendy vectors to convey phishing email attacks in the real world~\cite{proofpoint2024phish}. In this paper, we aim to understand these threats by focusing on their effectiveness against humans---the true target of phishing~\cite{yuan2024adversarial}.

Specifically, we focus on addressing two gaps in our knowledge---which, if filled, would enable current organizations to better cope with the never-ending struggle against phishers. Namely:
\begin{enumerate}[leftmargin=*]
    \item Quishing emails are gaining traction as a phishing vector~\cite{talos2024qr,proofpoint2024phish}. This raises a critical question: how effective are QR-codes at luring potential victims to a phishing website compared to ``traditional'' phishing vectors? Although quishing emails can evade automated detection mechanisms (§\ref{sssec:quishing}),  \textit{it remains unclear whether QR codes are equally effective at deceiving human users}.
    One might expect traditional phishing emails to perform better in this regard,\footnote{Clicking a link is nearly effortless, whereas scanning a QR code involves multiple steps: {\scriptsize \textit{(i)}} retrieving a secondary device such as a smartphone, {\scriptsize \textit{(ii)}} opening a QR-code reader, {\scriptsize \textit{(iii)}} scanning the code, and {\scriptsize \textit{(iv)}} following the link on the separate device.} as the additional burden introduced by QR codes could reduce user engagement.
    If this expectation does not hold, however, it would suggest an urgent need to adapt both {\small \textit{(i)}} automated detection systems and {\small \textit{(ii)}} phishing education programs to better address the rising threat of quishing.
    
    \item There is increasing evidence of LLMs being used by phishers in the wild~\cite{proofpoint2024phish,egress2024email}. This raises the question: how susceptible are a given company's employees against phishing emails written by an LLMs fed with OSINT information pertaining to their targeted company? Indeed, attackers are increasingly refining their tactics and can develop automated OSINT pipelines capable of targeting the entire workforce of a given organization. 
\end{enumerate}
Moreover, to provide an additional human-centered perspective, we attempt to establish whether employees ``perceived'' \textit{phishing awareness} has any relationship with phishing susceptibility to our considered phishing threats. Such an investigation is motivated by the many works that address the topic of phishing education~\cite{lain2024content, ho2025understanding, schiller2024employees}, which often show contrasting results (c.f.~\cite{lain2022phishing} with~\cite{ho2025understanding}).

We were unable to identify any prior work that specifically addresses the first gap.\footnote{We systematically review the (lack of) coverage of ``quishing'' in Appendix~\ref{app:systematic}.} 
For the second gap, the only real-world evidence we found comes from two  studies~\cite{ibm2023llm,bethany2024large}, each focused on a single organization---reflecting a broader trend in phishing research~\cite{lain2022phishing,schiller2024employees,burda2023peculiar,lain2024content}.
While these studies---like ours---do not claim universal generalizability, it remains unclear whether particular attack methodologies can be broadly applied across different organizational contexts. It is also of interest to explore how the same phishing approach might yield varying outcomes depending on the organizational environment. We therefore aim to investigate these knowledge gaps through a multi-organization study.

\section{Research Questions and Problem Definition}
\label{sec:rq}

\noindent
In this work, we tackle a broad research question (RQ): 
``\textit{How resilient employees across organizations of different size are to emerging social engineering techniques used to deliver phishing email attacks}?'' Specifically, to align such an RQ with the previously identified research gaps, we disentangle this RQ into three sub-RQs:
\begin{cooltextbox}
\begin{itemize}[leftmargin=0.55cm]
    \item[{\small RQ1}] Are Quishing emails more (or less) effective at deceiving end users than traditional button-based ``click-through'' emails?
    \item[{\small RQ2}] What are the effects of LLM-generated and OSINT-based phishing emails against modern organizations' employees?
    
    \item[{\small RQ3}] Is there a correlation between employees' {\small \textit{(a)}}~perceived phishing awareness and their {\small \textit{(b)}}~actual susceptibility to phishing?
    
\end{itemize}
\end{cooltextbox}

\noindent
To better understand the framing of our research questions, we now present the fundamental assumptions that drive our experiments.

\subsection{Threat Model} 
\label{ssec:threat}

\noindent
It is evident that, to address RQ1 and RQ2, we need to craft phishing emails and measure their effectiveness. Let us elucidate the quintessential security elements of the overarching scenario.

We assume an attacker that wants to steal sensitive credentials (i.e., userid and password) of employees of a given target company. The attacker seeks to do so via targeted phishing emails, which include elements mimicking those of the target company, which are sent to an unspecified set of employees of such a company. Therefore, we assume the attacker knows the email addresses of some employees as well as their name and surname (inferring the email given the name/surname, or vice versa, is easy~\cite{veluru2013mail,polakis2010using}). 
The attacker also has some knowledge on the target company, such as what provider is used to handle company-related emails (e.g., Microsoft or Google; inferring such information is doable, e.g., via MX lookups~\cite{deccio2021measuring}). 
To harvest credentials, the attacker sets up a malicious webpage that mimics the organization’s branding (e.g., logos, banners) to foster a sense of authenticity~\cite{williams2019persuasive,kim2021security}. In our experiments, we assume that the URL of the malicious site is not (yet) listed in any blocklist.\footnote{If the URL is included in some blocklists, then the phishing campaign would likely fail because any victim may not reach the credential-harvesting webpage---either because the webpage is blocked by the browser, or by a firewall; or even because the email may be blocked by the automatic filters and hence not read in the first place.} Practically, this can be achieved by deploying cloned versions of the webpage to different URLs~\cite{koide2023phishreplicant}. 
The attacker leverages their (limited) knowledge of the target company to craft a phishing email designed to lure recipients to a malicious webpage. To conceal the suspicious nature of the URL, the attacker may either embed it in a click-through button or encode it into a QR code included in the email body. RQ1 investigates the relative effectiveness of these two phishing tactics.

Moreover, we consider a scenario where the attacker leverages openly accessible LLMs to (cheaply) generate the phishing email. To this end, the attacker gathers publicly available information about the target company (e.g., from social media or the company's website) and provides it as input to the LLM. The model is then tasked with extracting relevant details and generating a persuasive email that could deceive recipients. RQ2 explores how OSINT can be combined with LLMs to craft phishing emails and evaluates the practical implications of such strategies.

\subsection{Experimental Approach and Choices}
\label{ssec:approach}

\noindent
We describe our approach, justifying two crucial design choices.

\textbf{Generic approach (and challenges).} First, we must find some organizations that enable us to collect data for our RQ. Specifically, these organizations must grant us the following permissions:
\begin{itemize}[leftmargin=*]
    \item Carry out phishing simulations (or give us data about phishing simulations) which entail a large share of their employees.
    \item Give us some freedom on such simulations, so that we can craft the ``phishing'' emails in such a way that we can answer RQ1--2.
    \item Enable some form of interactivity with some of their employees to measure the ``perceived'' phishing awareness for RQ3.
\end{itemize}
More specifically, for RQ1, we need to test the effectiveness of two emails---which should be identical, aside from: one leading to the credential-harvesting webpage via a click-through button; and another one via a QR-code. For RQ2, we need a third email created by collecting OSINT on the company (hence, even though OSINT entails publicly available information, we must still obtain the company's permission to carry out OSINT activities). Nevertheless, to ensure consistency in the data we collect (and hence provide a meaningful answer to our RQs), we must design our experiments by minimizing the underlying differences that exist between the companies that accept to collaborate in this research.
In what follows, we provide more details on how we seek to answer our RQs.

\textbf{Phishing email effectiveness (and susceptibility).} Our research is centered on a core objective: measuring the impact of phishing \textit{emails' content} to deceive users. In simple terms, we want to measure the following: ``given a (phishing) email that is \textit{read} by a user, will such a user be fooled and hence \textit{visit} the (malicious) webpage pointed by the email?'' We do not consider emails that have not been opened, since such an outcome is not related to the email's content---but rather to its metadata or external factors (e.g., subject or time of delivery); at the same time, what happens after the user visits the landing webpage can be influenced by other factors, such as browser, device, or even the landing page itself---all of which have little relevance to the email's content. Therefore, we measure the effectiveness of a set of phishing emails (or the susceptibility of a company's employees to a set of phishing emails) by computing the ratio of those phishing emails that successfully bring a user to the corresponding landing webpage with respect to the number of phishing emails that have been read: a higher ratio denotes more effective emails (or more susceptible employees). Such a metric is typical in related studies~\cite{petelka2019put, siadati2017measuring,hillman2023evaluating,heiding2024evaluating,sarno2022so}.

\textbf{Measuring the PPA.} There are various ways (e.g.,~\cite{chaudhary2022developing, manifavas2014dsape, rantos2012effective, aba2016perceived}) to collect data that can be used to measure the awareness of a given set of subjects with respect to phishing threats---which is a core theme in cybersecurity awareness programs (CSA). For our research, we focus on the perception of the end users, i.e., the ``perceived'' phishing awareness (PPA). We do so by following the guidelines by Chaudhary et al.~\cite{chaudhary2022developing}, who indicate \textit{surveys} as the most appropriate (and least intrusive) mechanism to collect data useful for our goal.\footnote{Chaudhary et al.~\cite{chaudhary2022developing} synthesized 32 papers on cybersecurity awareness programs, and proposed four indicators: \textit{impact} (which is our focus, given that its relatedness to phishing susceptibility), \textit{sustainability}, \textit{accessibility}, \textit{monitoring}. Each indicator comprises nine factors: \textit{attitude towards cybersecurity}, \textit{interest}, \textit{usability}, \textit{self-reported behavior}, \textit{knowledge and competence gain}, \textit{value added}, \textit{reachability}, \textit{touchability}, \textit{overall feedback}.} Building on the foundations of prior work, our survey is rooted on the ``knowledge-attitude-behavior'' principles~\cite{schrader2004knowledge}: \textit{knowledge} denotes ``familiarity, awareness, or understanding'' of security policies/procedures/standards/directives/regulations/laws/guidelines/strategies/technologies. \textit{attitude} denotes ``beliefs/opinions/thinking/feelings'' toward security; \textit{behavior} denotes the way a person ``acts'' when faced with security issues~\cite{chaudhary2022developing}. In this context, RQ3 seeks to correlate information related to the PPA of a given company's (subset of) employees with the overall susceptibility of such a company's employees to phishing emails---and then see if the same result holds across different companies.
\section{Methodology}
\label{sec:method}
\noindent
To systematically answer our RQs, we found agreements with three companies (described in~§\ref{ssec:companies}).  We first carried out three phishing simulations in these companies (described in~§\ref{ssec:setup}), focused on answering RQ1 and RQ2. Afterwards, we focused on RQ3 and carried out a user study (described in §\ref{ssec:awareness_implementation}) with our companies' employees. We discuss ethical concerns pertaining to our methodology in §\ref{ssec:ethics}.

\subsection{Description of Companies}
\label{ssec:companies}

\noindent
We contacted various companies located in Central Europe, asking for their collaboration in a research project on the topic of phishing email susceptibility and assessment. After numerous exchanges, we eventually found an agreement with three companies, which we summarize below (an overview is provided in Table~\ref{tab:companies}).
\begin{itemize}[leftmargin=*]
    \item \textit{Small-sized Company} (\cs{}). A small-medium enterprise (SME) operating in the hospitality sector~\cite{kansakar2019technology}. \cs{} has between 50 and 250 employees. The customers of \cs{} are located only in two countries in Europe. This company has no self-administered IT department, and their IT is outsourced to a third-party vendor. In terms of CSA training, only basic dissemination methods (e.g., educational texts and slides covering various attack vectors seen in the past) are adopted by \cs{}, which are provided to its employees on a yearly basis. Each employee receives the exact same CSA training, and there is no mechanism meant to assess the effectiveness of CSA training in \cs{}. Notably, \cs{} had never carried out in-house phishing simulations before.

    \item \textit{Medium-sized Company} (\cm{}). An enterprise with less than 2\,000 employees, operating in the financial sector. This company operates in multiple German-speaking countries as well as in some middle-eastern countries. The employees of \cm{} span across the typical roles of financial organizations, including administration, marketing, and IT. Indeed, \cm{} has its own IT infrastructure and a dedicated cyber security team, which is also tasked to carry out phishing simulations at least twice per year. The CSA training provided by \cm{} to its employees includes slides, videos, texts, as well as physical classes covering even recent/emerging phishing trends. Each employee receives the same type of CSA training (on a yearly basis) and employees are required to pass a dedicated exam to demonstrate their understanding of such CSA training. 

    \item \textit{Huge-sized Company} (\ch{}). A multi-national enterprise (having over 30\,000 employees) conducting business on a global scale in the manufacturing sector. They have a sophisticated IT infrastructure and a large cybersecurity team. \ch{} regularly carries out phishing simulations across all its employees. CSA training is provided every two years, and is tailored for the specific role of each employee (e.g., managers receive different training than members of HR). Such training includes slides, videos, texts, in-person training, and even leverages eLearning platforms with additional content (covering also recent trends); at the end of each CSA training campaign, employees must pass an exam.
\end{itemize}
Information on the CSA of each company has been derived via a structured questionnaire (Table~\ref{tab:questionnaire_2} in Appendix~\ref{app:questionnaires}) with knowledgeable representatives of each company. Due to non-disclosure agreements (NDA), we cannot provide more details.

\begin{table}[!t]
    \centering
    \caption{\textbf{Overview of Companies.} \textmd{For our research, we considered three companies whose businesses is predominantly located in Central Europe.}}
    \vspace{-3mm}
    \resizebox{0.99\columnwidth}{!}{
        \begin{tabular}{@{ }l|c|c|c@{ }}
            \toprule
             & \textbf{Small Company} (\cs{}) & \textbf{Medium Company} (\cm{}) & \textbf{Huge Company} (\ch{})\\
            \midrule
            \textit{\# Employees} & between 50 and 250 & \smamath{\approx}1\,500 & \smamath{>}30\,000 \\
            \textit{Industry} & Hospitality & Finance & Manufacturing \\
            \textit{CSA Training Frequency} & Yearly & Yearly & Biyearly \\
            \textit{CSA Training Approaches} & Slides, Texts & Slides, Videos, Texts, Classes & Slides, Videos, Text, Classes, eLearning \\
            \textit{In-house Simulations?} & \xmark{} & \cmark{} & \cmark{} \\
            \textit{CSA Training Specificity} & Generic & Generic & Group-specific \\
            \textit{Emerging Trends in CSA?} & \xmark{} & \xmark{} & \cmark{} \\
            \textit{Simulation Framework} & (GoPhish~\cite{gophish}) & MS Defender~\cite{defender} & MS Defender~\cite{defender} \\
            \bottomrule
        \end{tabular}
    }
    \label{tab:companies}
    \vspace{-3mm}
\end{table}

\begin{figure*}[!htbp]
\centering
    \begin{subfigure}[!htbp]{0.65\columnwidth}
        \centering
    \frame{\includegraphics[height=6.7cm]{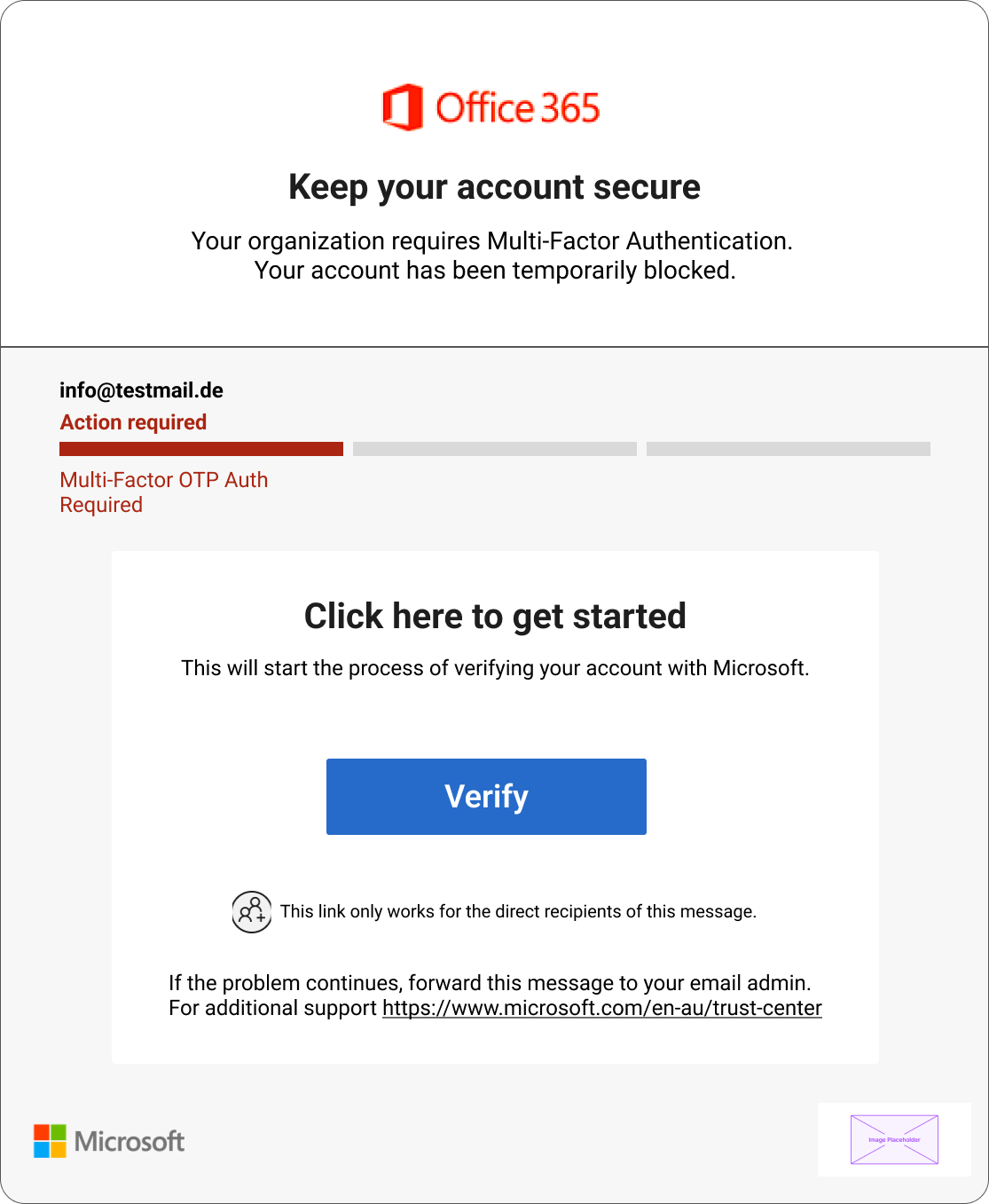}}
    \caption{Example of button ``click-through'' email (\eb[\scriptsize]). \textmd{The ``info@testmail.de'' was replaced with a company-related email address.}}
    \label{sfig:email1}
    \end{subfigure} 
    \hspace{1mm}
    \begin{subfigure}[!htbp]{0.65\columnwidth}
        \centering
    \frame{\includegraphics[height=6.7cm]{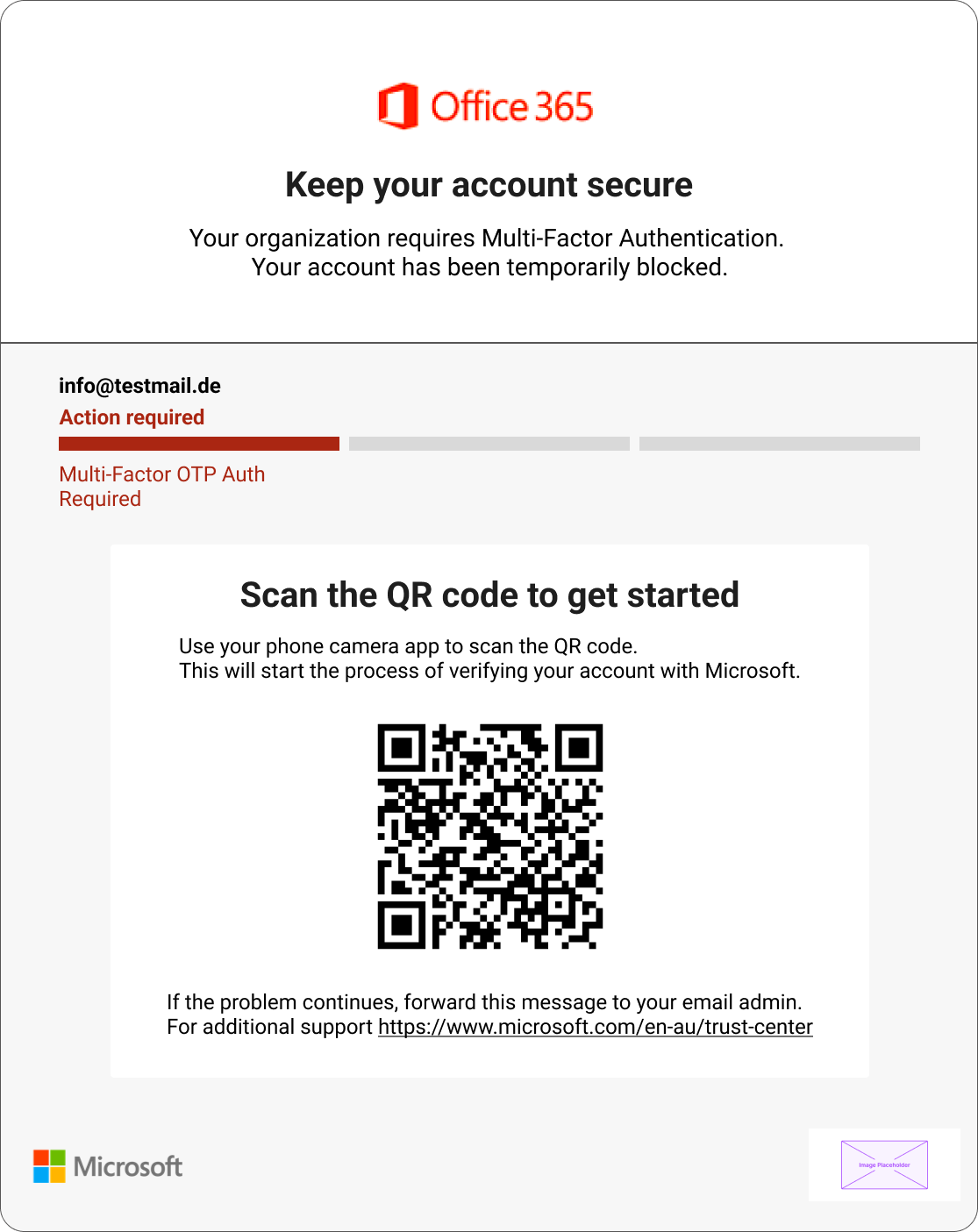}}
    \caption{Example of QR-code phishing email (\eq[\scriptsize]). \textmd{Note that the design is identical to \eb[\scriptsize] aside from the button being replaced with a QR-code.}}
    \label{sfig:email2}
    \end{subfigure} 
    \hspace{1mm}
    \begin{subfigure}[!htbp]{0.65\columnwidth}
        \centering
    \frame{\includegraphics[height=6.7cm]{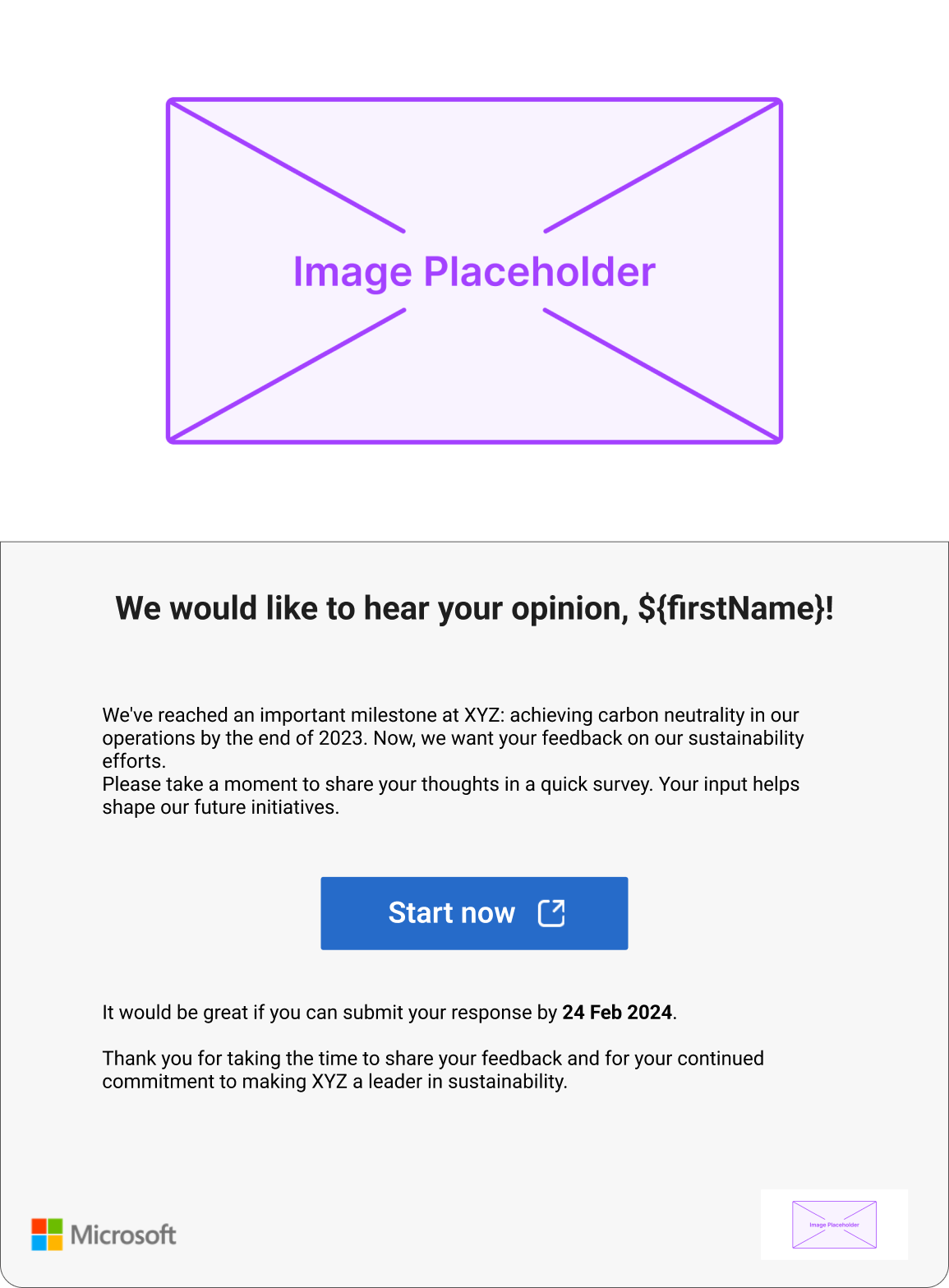}}
    \caption{Example of OSINT+LLM phishing email (\el[\scriptsize]). \textmd{The large ``image placeholder'' was replaced with an image taken from a press release of the specific company.}}
    \label{sfig:email3}
    \end{subfigure}

    \vspace{-3mm}
    \caption{\textbf{Emails used in our experiments.} \textmd{Our emails shared a similar design, but each email presented some company-specific traits to increase authenticity (e.g., we put the company logo at the bottom right). All emails bring the user to the same landing webpage (which was also specific to each company).}}
    \label{fig:emails}
    \vspace{-4mm}
\end{figure*}

\subsection{Experimental Setup of the Simulations}
\label{ssec:setup}

\noindent
We create three ``phishing'' emails: two (described in §\ref{sssec:simulation_qr}) for RQ1, whereas a third (described in §\ref{sssec:simulation_ai}) is for RQ2. Let us first introduce the common elements of our experimental testbed~(§\ref{sssec:common}).

\subsubsection{\textbf{Common elements\\}}
\label{sssec:common}
All phishing simulations carried out in our study have been designed in accordance with the chosen companies' policies. Specifically, our simulations were meant to serve as a periodical assessment of \cm{} and \ch{}; whereas, for \cs{}, our simulations were the very first phishing assessment done in \cs{}. Such a context allowed us to develop a customized simulation framework for \cs{}, for which we aligned with \cm{} and \ch{} to minimize inter-company differences. In what follows, we provide the most relevant technical details.
\begin{itemize}[leftmargin=*]
    \item \textit{Infrastructure.} At a high-level, our experiments entail  sending ``phishing'' emails to employees and see how they react (e.g., whether they open the email, or click on the phishing link). We used the existing infrastructure (i.e., Microsoft Defender~\cite{defender}) for \cm{} and \ch{}. For \cs{}, we deployed our own infrastructure by leveraging the open-source GoPhish framework~\cite{gophish} (used also, e.g., in~\cite{burda2023peculiar, lokesh2023data}). Altogether, these frameworks enable collection of the data required for our RQs (see §\ref{ssec:approach}). We provide low-level details on the experimental infrastructure, as well as on the challenges we had to overcome to set it up, in Appendix~\ref{app:technical}.

    \item \textit{Landing webpage}. 
    A common practice~\cite{lain2022phishing} in phishing simulations is to embed a link in the email that points to a ``credential-harvesting'' webpage that invites the user to submit some sensitive information---which aligns with our threat model~(§\ref{ssec:threat}). Such a webpage is typically designed to enable tracking of those users that land on it, or who submit their data. Both \cm{} and \ch{} used the typical login page of Microsoft for their simulations, so we designed the landing page for \cs{} (shown in Fig.~\ref{fig:landing} in the Appendix) accordingly, given that \cs{} also relies on Microsoft.\footnote{The landing webpage for \cs[\scriptsize] was hosted on a domain we purchased ourselves, and we made the URL very similar to that of the official webpage of \cs[\scriptsize]. Specifically, its URL was ``\$Company$_{s}$Name.email''. The landing webpages for \ch[\scriptsize] and \cm[\scriptsize] were hosted on their premises; we cannot provide details on these URL due to NDA.} 

    \item \textit{Data collection.} We coordinated with the companies so that each (randomly chosen) employee could only receive at most one email among those we crafted. We timed the delivery so that each recipient would receive the email in the morning (around 8am, accounting for timezones) of a work day. All emails include a ``Microsoft'' component (e.g., a logo) since all companies use the Microsoft Office suite. The sender of these emails was always related to an identity resolving to ``@mircosoft.com''. When crafting our emails, since they entailed HTML objects (e.g., images), we ensured that they rendered correctly on the email clients most commonly used by the respective company's employees.\footnote{While setting up our testbed for \cs[\scriptsize] we noticed that their default configuration of Microsoft Outlook forced the user to explicitly allow displaying images (including QR codes) in an email before showing them---if the email comes from an ``unknown'' sender. To overcome this problem, the sender of our emails was added to the ``trusted sender'' list. This ensured that any images (including the QR code) would be displayed---thereby also guaranteeing a correct counting of the opened emails by GoPhish  (see~\cite{burda2020don}).} Finally, for each email sent, we obtained: the number of recipients that opened/read it; the number of recipients that visited the landing webpage; the number of recipients that reported the message, and the number of submitted credentials.
    
\end{itemize}
The simulations occurred between April 24th and May 10th, 2024 (depending on the company), and lasted around 3 days each. 

\subsubsection{\textbf{Quishing \& Traditional-phishing Emails}\\}
\label{sssec:simulation_qr}

To answer RQ1, we carried out two phishing simulations, each revolving around a specific email created ad-hoc for our experiments. 

\begin{itemize}[leftmargin=*]
    \item \textit{Button email} (\eb{}). This email contains an URL integrated in a ``click-through'' button that brings the user to the landing page.

    \item \textit{QR-code email} (\eq{}). This email is identical to \eb{}, and the only difference is that \eq{} includes a QR code (instead of the button).
\end{itemize}
Fig.~\ref{sfig:email1} shows an example of \eb{}, while Fig.~\ref{sfig:email2} presents its quishing counterpart, \eq{}. The two emails are visually identical, except for the interaction mechanism: \eb{} prompts users to click a button, whereas \eq{} requires scanning a QR code (e.g., via smartphone) to access the concealed URL. This controlled design isolates the variable of interest, enabling us to address our first RQ.

\textbf{Email design.} To create the ``phishing hook'' of these emails, we took inspiration from the common tactics adopted by phishers. Specifically, the email urges the recipient to set up multi-factor authentication to reactivate their account---which should be done by following the link included in the email. We chose such a hook because it is well-known (e.g.,~\cite{gordon2019assessment}) that emails containing IT-related topics are very successful at deceiving users. Moreover, we created the emails so that they had an ``authentic'' design~\cite{aljeaid2020assessment}, resembling that of communications sent by the respective company (i.e., we used the company's logo and also elements of the Microsoft Office suite). We also included urgency cues (e.g., ``action required'') and loss (``your account has been temporarily blocked''), since they have all been found to be very effective~\cite{williams2018exploring}. Note that all of our design choices comply with the overarching threat model~(§\ref{ssec:threat}).

\textbf{Company-specific differences.} 
Full alignment across companies was not always feasible, so minor differences affect our setup.
First, while the emails were in English for \ch{}, they were translated into German for \cs{} and \cm{}, where German is the primary language. Second, a security banner (``This is an external email...'') was included for \cm{} and \ch{}, but not for \cs{}, which does not use such warnings by default. Third, \ch{} did not require \eq{}, as it had recently conducted a QR-based simulation (in Jan. 2024), for which it shared results with us. Accordingly, we designed \eb{} to match that version of \eq{} for \ch{}, differing only in the use of a button instead of a QR code; we cannot disclose the exact emails due to NDA. Importantly, these variations do not affect our investigation of RQ1, as \eq{} and \eb{} retain consistent properties within each organization. 

\textbf{Data analysis.} Overall, \smamath{18\,339} \eb{} (\smamath{21} for \cs{}, \smamath{567} for \cm{}, \smamath{17\,751} for \ch{}) and \smamath{34\,610} \eq{} (\smamath{21} for \cs{}, \smamath{558} for \cm{}, \smamath{34\,031} for \ch{}) were sent.\footnote{For \ch[\scriptsize], the numbers for \eq[\scriptsize] are higher than \eb[\scriptsize] because the quishing simulation in \ch[\scriptsize] had been carried out by \ch[\scriptsize], and it hence targeted all employees that take part in these phishing simulations; in contrast, the simulation for \eb[\scriptsize] has been carried out by us, and the emails were sent by randomly choosing half of the employees of \ch[\scriptsize] (the remaining half received the email used for RQ2, discussed in §\ref{sssec:simulation_ai}).
Across all companies, our emails were addressed to approximately 30–50\% of employees. Due to NDA restrictions, we cannot disclose the exact percentages.}
As we explained (§\ref{ssec:approach}), the answer to our first RQ is determined by calculating, for both \eq{} and \eb{}, the ratio of those users that visited the landing webpage with respect to those that read the email; and then compare these two numbers via statistical tests and analytics.

\subsubsection{\textbf{Phishing Email by using OSINT and LLM\\}}
\label{sssec:simulation_ai}

\begin{figure}[!t]
    \centering
    \includegraphics[width=0.9\columnwidth]{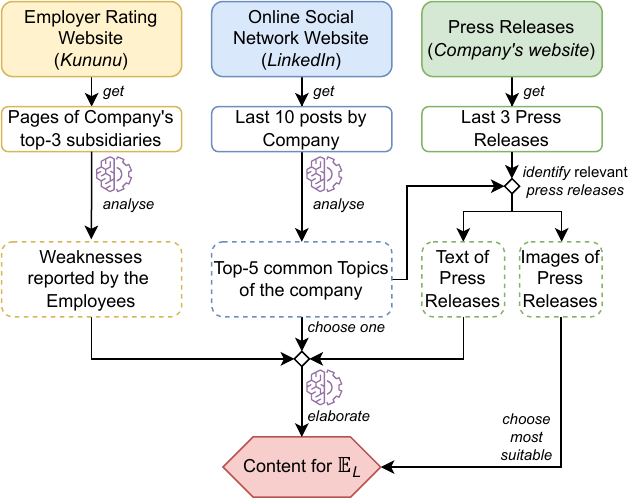}
    \vspace{-3mm}
    \caption{\textbf{Extraction and exploitation of OSINT for \el[\scriptsize].} \textmd{Operations denoted with a ``brain-cog'' image have been carried out with an LLM.}}
    \label{fig:osint}
    \vspace{-4mm}
\end{figure}

\noindent
To address RQ2, we carry out a simulation revolving around a single email, \el{}, crafted by providing OSINT-acquired information as input to a publicly accessible (and free to use) LLM. At a high level, our methodology resembles that used in the test by X-Force in 2023~\cite{ibm2023llm}. Specifically, our email aimed to imitate the invitation to participate in survey, organized by the targeted company, on topics aligning with the company's agenda. To create \el{}, we followed a systematic approach (visualized in Fig.~\ref{fig:osint}), rooted on the assumptions of our threat model~(see §\ref{ssec:threat}). We describe it below.

\textbf{Gathering OSINT.}
We leveraged three publicly available sources: the most prominent German-speaking employer rating website (Kununu~\cite{kununu}), an online professional social network (LinkedIn~\cite{linkedin}), and press releases on the company. We chose these sources because of their popularity and relevance---given the considered companies' public visibility. Let us explain how we used each of these sources.

\begin{itemize}[leftmargin=*]
    \item \textit{Kununu.} We began by surveying the Kununu page for each company, seeking to find ``issues/weaknesses'' reported by the employees. We found that \cs{} did not have a dedicated entry on this website. For \cm{} and \ch{}, we first identified the three largest subsidiaries of each company; then, we looked up the Kununu page of each subsidiary; finally, we saved the Kununu's page (if available) of each subsidiary in a single HTML file---which would be later used as input to the LLM to find ``potential weaknesses'' that could be exploited for phishing attacks.

    \item \textit{LinkedIn.} Afterwards, we looked at the LinkedIn page of each company. The goal was finding a topic that could be of interest to the employees of the company---the recipients of our email. All companies have an official page on LinkedIn. We selected the 10 last posts made by the company (excluding reposts), and saved the contents of such posts into a single HTML file. Such a file would be later used as input to the LLM to (automatically) find ``the 5 most common topics covered in these posts''.

    \item \textit{Press releases.} Finally, we considered the public press releases of each company, for which we relied on the official communication channels in each company's website. The idea was to further enrich our previous analyses by also integrating visual assets that could improve the quality of our phishing email. Hence, for each company, we considered the three most recent press releases: after identifying those that deal with the topics found via LinkedIn, we downloaded all usable assets (e.g., images or banners) and we saved the text of the press release in a file---which we later used as input to the LLM to write ``an introductory text to a survey about a relevant topic for this company''.
    
\end{itemize}   
The above-mentioned operations were done in February 2024.

\smallskip
\textbf{Feeding OSINT to the LLM.} 
We considered ChatGPT 3.5 Turbo for our LLM. Our choice is because it was free: ChatGPT 4.0 required a paid subscription (at the time) which may have discouraged real attackers from using it (phishing campaigns tend to be cheap~\cite{spacephish2022}). Nonetheless, to generate the content of each email, we assembled a sequence of five prompts (see Table~\ref{tab:prompts} in Appendix~\ref{app:technical}) that integrate the information extracted via our OSINT operations. 
The resulting text was used to compose the main body of our email \el{}. Then, we used the most appropriate visual assets we collected from the press releases and used them to improve the aesthetics of the email. Next, to provide a sense of authenticity, we added the company's and Microsoft's logos at the bottom of the email. Finally, we added a click-through button that embedded a link to our landing page (ideally, the survey was meant to be organized by the company so that only its employees could participate---which is why a login was required) and added text soliciting the recipient to submit their responses within a few days. To sum up, we used the LLM to both {\small \textit{(i)}}~summarize content and {\small \textit{(ii)}}~write the email's text.

\textbf{Data analysis.} Overall, \smamath{18\,360} \el{} (\smamath{18} for \cs{}, \smamath{589} for \cm{}, \smamath{17\,753} for \ch{}) have been sent. To investigate RQ2, we qualitatively analyse all numerical data we can collect related to \el{}. We predominantly focus on the ratio of users that visited the landing webpage w.r.t. those that read the email; but we also gauge the ratio of users that submitted their credentials w.r.t. those that visited the landing webpage. This is because, for this email, we do not leverage the sense of loss as we did for \eb{} and \eq{} (i.e., ``your account has been temporarily blocked''). Thus, users may not expect to land on a webpage that asks them to input their credentials. Therefore---contrarily to \eb{} and \eq{}---for \el{} the submission of credentials is strongly dependent on how persuasive \el{}'s content (which depends on OSINT and LLM, i.e., the crux of RQ2) is in deceiving the end user.

\subsection{Perceived Phishing Awareness}
\label{ssec:awareness_implementation}

\noindent
To answer RQ3, we conducted user surveys among companies' employees to estimate their (perceived) phishing awareness.
The surveys (implemented via MS Forms) consist of anonymous closed-answer questionnaires, developed in agreement with each company.

\textbf{Questionnaire.}
The questionnaire follows scientific guidelines on empirical social research~\cite{baur2014handbuch}, and is rooted on the knowledge-attitude-behavior principles~\cite{schrader2004knowledge}. As we explained (§\ref{ssec:approach}), our questionnaire is built on the work of Chaudhary et al.~\cite{chaudhary2022developing}. Specifically, to allow complete coverage of the ``impact'' indicator (which is related to phishing susceptibility), we consider the following factors proposed by Chaudhary et al: \textit{attitude towards cybersecurity}, \textit{interest}, \textit{usability}, \textit{behavior}, \textit{knowledge and competence gain}.\footnote{N.b.: our preliminary survey with the companies' representatives (shown in Table~\ref{tab:questionnaire_2}), used to derive the CSA profile of each company (summarized in §\ref{ssec:companies}), covered the remaining four factors (i.e., \textit{value added}, \textit{reachability}, \textit{touchability}, \textit{overall feedback}). Hence, our study provides a complete coverage of the ``impact'' indicator.} 
Overall, the questionnaire spans across 40 questions, distributed in five sections: 
{\small \textit{(i)}}~attitude towards cybersecurity---for which we provide a snippet in Fig.~\ref{fig:survey_attitude}; 
{\small \textit{(ii)}}~cybersecurity routines---which focuses on the \textit{(self-reported) behavior}; 
{\small \textit{(iii)}}~cybersecurity awareness training---which focuses on \textit{CSA training experience} (related to \textit{interest}) and \textit{training usability};
{\small \textit{(iv)}}~quick assessment---for which we provide a snippet in Fig.~\ref{fig:survey_quick}, and which focuses on \textit{knowledge and competence gain};
{\small \textit{(v)}}~socio demographics. The answers to most of the questions were based on a 5-point Likert Scale~\cite{norman2010likert} (1=strongly disagree; 5=strongly agree).
Given the multi-lingual nature of our companies, the questionnaire was created both in German and in English. One of the authors has native German fluency. The complete list of questions in our questionnaire is provided in Table~\ref{tab:questionnaire_1} (in Appendix~\ref{app:questionnaires}).

\textbf{Distribution.}
For \cs{} and \cm{}, our questionnaire was distributed among employees via the official internal communication platform; for \ch{}, it was distributed via dedicated ``team'' channels as well as by convenience sampling~\cite{emerson2015convenience} (e.g., by sending emails to employees). 
For each company, we disseminated the questionnaire a few days after concluding our phishing simulations and collected responses for approximately one week.
While we could not control exactly who chose to participate, it is reasonable to expect that respondents had also taken part in the phishing simulation.

\textbf{Data analysis.} To answer RQ3, we first scrutinize the data we collected with our questionnaire, and then cross-analyse our results with those of our phishing simulations. Specifically, for each company, we first compute the mean and variance of the five-point Likert scale for each item of our questionnaire; then, we derive a general ``Perceived Phishing Awareness'' score (PPA-score) by aggregating all responses. 
Next, we define a ``Phishing Susceptibility'' score (PS-score) by computing the ratio of those emails (accounting for \eb{}, \eq{}, and also \el{}) that brought a user to a landing webpage w.r.t. those that have been read. 
Considering a single pool which aggregates the results of all our emails (i.e., \smamath{71\,309} in total) is valid because, despite some differences: the emails are ultimately all ``phishing'', leveraged the same structural properties, and there was no targeted or arbitrarily-chosen selection of recipients (any employee could be eligible).\footnote{Moreover, given that we do not know which email (among \eb[\scriptsize], \eq[\scriptsize], \el[\scriptsize]) was received by our participants, it would be unfair to consider the results of a single simulation.} Finally, we statistically compare the PPA-score with the PS-score, and draw our conclusions.

\subsection{Ethical Considerations}
\label{ssec:ethics}
\noindent
To ensure we perform our simulations ethically, we followed well-known and established scientific practices~\cite{bailey2012menlo,kohno2023ethical,baur2014handbuch}. 

Our experiments (§\ref{ssec:setup}) have been designed in accordance with company \ch{} and \cm{}'s ethical standards, whereas \cs{} subcontracted us to carry out the assessment within their premises. 
Accordingly, all activities were approved by the relevant ethical bodies within the involved organizations. 
Employees that received one of our phishing emails were aware that their companies carried out phishing assessments. 
No harm was caused during the study: both the emails and landing pages were under our control, and no credentials were persistently stored.
Participants of our (anonymous) user study (§\ref{ssec:awareness_implementation}) did so willingly and had been made aware that their responses would be collected and used only for scientific purposes. 

Finally, we mention that ChatGPT refused to answer our request to provide ``weaknesses that could be leveraged for a phishing attack'' (§\ref{sssec:simulation_ai}), so we had to find a workaround that would bypass its automatic censorship mechanisms (we do not provide our exact prompts to avoid helping attackers crafting phishing emails).

\section{Phishing Simulations [RQ1, RQ2]}
\label{sec:simulations}

\noindent
We first present the results (§\ref{ssec:simulations_results}) of our three simulations entailing the QR-code phishing email (\eq{}), the traditional phishing email leveraging a click-through button (\eb{}), and the OSINT-fed LLM-written email (\el{}). Then, we address our first research question with a statistical test (§\ref{ssec:rq1}). Finally, we address our second research question (§\ref{ssec:rq2}) with a comprehensive qualitative analysis.

\subsection{Overall Results of our Phishing Simulations}
\label{ssec:simulations_results}

\noindent
We report in Table~\ref{tab:simulations_results} the results of our simulations. Specifically, for each email (\eb{}, \eq{}, \el{}) we provide: the overall number of emails sent; the number of emails that have been read (we consider an email as ``read'' if it has been opened); the number of emails for which the landing page has been visited at least once; the number of emails for which the credentials of the recipient have been submitted; as well as the ``page visited / email read'' ratio and the ``credential submitted / email read'' ratio. The results are provided for each company, and the rightmost columns report the aggregated results across all companies. 
Let us analyse our results at a high level.

\begin{table*}[!t]
    \centering
    \caption{\textbf{Results of \eb[\scriptsize], \eq[\scriptsize], and \el[\scriptsize].} \textmd{We recall (§\ref{sssec:simulation_qr}) that, for \ch[\scriptsize], the simulation of \eq[\scriptsize] was not managed by us: the email was sent to more employees and no data was logged about the credentials submitted. Therefore, numbers with an asterisk~(*) have been derived by removing the \eq[\scriptsize] of \ch[\scriptsize] from the pool.}}
    \vspace{-3.5mm}
    \resizebox{1.8\columnwidth}{!}{
    \setlength{\tabcolsep}{6.2pt}
        \begin{tabular}{@{ }l?c|c|c||c|c|c||c|c|c?c|c|c@{ }}
            \bottomrule
            \textbf{Company} & \multicolumn{3}{c||}{\cs{} {\small (Small Company)}} & \multicolumn{3}{c||}{\cm{} {\small (Medium Company)}} & \multicolumn{3}{c?}{\ch{} {\small (Huge Company)}} & \multicolumn{3}{c}{AGGREGATE}\\
            \cmidrule{1-13}
             \textbf{Email} & \eb{} & \eq{} & \el{} & \eb{} & \eq{} & \el{} & \eb{} & \eq{} & \el{} & \eb{} & \eq{} & \el{} \\
            \toprule
            
            Emails sent & $21$ & $21$ & $18$ & $567$ & $558$ & $589$ & $17\,751$ & $34\,031$ & $17\,753$ & $18\,339$ & $34\,610$ & $18\,360$\\ 
            Emails read & $9$ & $13$ & $12$ & $312$ & $317$ & $397$ & $11\,538$ & $24\,842$ & $11\,025$ & $11\,859$ & $25\,172$ & $11\,434$ \\
            Page visited & $2$ & $3$ & $8$ & $12$ & $17$ & $125$ & $936$ & $1\,950$ & $499$ & $950$ & $1\,970$ & $632$ \\
            Credentials submitted & $1$ & $1$ & $3$ & $9$ & $6$ & $59$ & $531$ & n/a & $243$ & $541$ & $7$* & $305$ \\
            Page visited / Email read & $22.2\%$ & $23.1\%$ & $66.6\%$ & $3.9\%$ & $5.4\%$ & $31.5\%$ & $8.1\%$ & $7.9\%$ & $4.5\%$ & $8.0\%$ & $7.8\%$ & $5.5\%$ \\
            Cred. sub. / Email read & $11.1\%$ & $7.7\%$ & $25.0\%$ & $2.9\%$ & $1.9\%$ & $14.9\%$ & $4.6\%$ & n/a & $2.2\%$ & $4.6\%$ & $2.1\%$* & $2.7\%$ \\

            \bottomrule
        \end{tabular}
    }
    \label{tab:simulations_results}
    \vspace{-4mm}
\end{table*}

For \cs{}, we can observe that \eb{} and \eq{} led to a similar outcome in terms of ``page visited / email read''. We also find it worrying that \smamath{11.1\%} (and \smamath{7.7\%}) of those who read \eb{} (and \eq{}) eventually submitted their credentials; even more worrying is the effectiveness of \el{}. However, the relatively-small sample size for \cs{} prevents one from drawing sound conclusions from these numbers. Nevertheless, after we launched our simulations, we were contacted by some representatives of \cs{} in charge of IT matters: they reported that they had been messaged/called by 11 employees, asking about the legitimacy of the emails they had just received.

For \cm{}, we also see (as for \cs{}) that the trends for \eb{} and \eq{} are similar, with \smamath{3.9\%} (resp. \smamath{5.4\%}) of those that opened \eb{} (resp. \eq{}) visiting the landing page. Moreover (and also in line with \cs{}), \el{} seems to have been more effective than both \eb{} and \eq{}. Finally, we mention that, for \eb{}/\eq{}/\el{}, \smamath{241}/\smamath{155}/\smamath{182} employees reported the email, whereas \smamath{352}/\smamath{377}/\smamath{312} deleted it.\footnote{These results are not available for \cs[\scriptsize] but are available for \ch[\scriptsize], because GoPhish does not provide such a functionality---which is, however, integrated in Microsoft Defender.}

For \ch{}, the outcome of \eb{} and \eq{} are also somewhat similar, with \smamath{8.1\%} (resp. \smamath{7.9\%}) of those who opened \eb{} (resp. \eq{}) visiting the landing page. However (and differently from \cs{} and \cm{}) the impact of \el{} had a lower impact than both \eb{} and \eq{} (despite \el{} having been read by comparatively the same amount of recipients as both \eb{} and \eq{}). 
Finally, we mention that, for \eb{}/\eq{}/\el{}, \smamath{3\,039}/\smamath{10\,268}/\smamath{1\,637} employees reported the email, whereas \smamath{6\,567} deleted \eb{} and \smamath{6\,375} deleted \el{} (we do not have such data for \eq{}).

\subsection{Statistical Assessment of \eb[\large] and \eq[\large] [RQ1]}
\label{ssec:rq1}

\noindent
To objectively answer RQ1, we rely on one-tailed chi-square  tests~\cite{rana2015chi}. 

We define our core hypothesis as follows: ``An employee opening \eb{} is \textit{more likely} to visit the landing webpage than an employee opening \eq{}.'' 
Indeed, our expectation is that QR-codes are less effective than click-through buttons (see §\ref{sssec:quishing}), since a button simply needs to be clicked/tapped, whereas a QR-code must be scanned first. Such a procedure can be cumbersome, and some employees may think twice before doing so, potentially leading to postponement or forgetfulness of the task; an employee may even realize that the email is suspicious and not proceed at all. We stress, as we stated (§\ref{sssec:simulation_qr}), that we seek to measure whether there is any statistically significant difference in the ability of a QR code to bring a potential victim to a phishing webpage w.r.t. a traditional click-through button (note that both cases conceal the URL). What happens ``after'' the user lands on such a webpage is outside the scope of RQ1 (and, hence, of our null hypothesis for this test).

We perform the chi-square test four times: first on the aggregated data from all companies (yielding a larger sample size), and then separately for each company. This procedure is statistically valid, as the measured phenomenon is consistent across companies.
\begin{itemize}[leftmargin=*]
    \item \textit{Aggregate.} Therefore, we aggregate the results for \eb{} and \eq{} across our three companies. For \eb{}, \smamath{11\,859} employees opened it, and \smamath{950} (\smamath{8.01\%}) visited the landing page; whereas \smamath{10\,909} did not visit the landing page despite opening \eb{}. For \eq{}, \smamath{25\,172} employees opened it, and \smamath{1\,970} (\smamath{8.49\%}) visited the landing page, whereas \smamath{23\,202} did not visit the landing webpage despite opening \eq{}. The result of the test is a chi-square statistic of \smamath{0.353}. The corresponding (one-tailed) \smamath{p}-value is \smamath{.276}, indicating no statistically significant difference (assuming a significance level of \smamath{.05}). Therefore, this test indicates that our hypothesis cannot be accepted. Moreover, the effect size is small (\smamath{0.0031}), further confirming that any difference between \eb{} and \eq{} is negligible.\footnote{The difference in click-through rates of \eb[\scriptsize] and \eq[\scriptsize] is \scmath{0.18\%}, with a \scmath{95\%} confidence interval of \scmath{(-0.41\%, 0.78\%) \in \pm1\%}, confirming no statistically-significant difference.} 
    
    \item \textit{Company-specific.} We repeat the chi-square test for each company to verify if our findings hold even in specific contexts; for simplicity, we only report the results. For \cs{}, chi-square=\smamath{0.0}, one-tailed \smamath{p}-value=\smamath{1.0}, effect size=\smamath{0.0}. For \cm{}, chi-square=\smamath{0.514}, one-tailed \smamath{p}-value=\smamath{1.0}, effect size=\smamath{0.029}. For \ch{}, chi-square=\smamath{0.709}, one-tailed \smamath{p}-value=\smamath{.2}, effect size=\smamath{0.004}. Hence, our hypothesis cannot be accepted for each of these tests. Note that, for \cs{}, the sample size is small (so it is possible that the test is inconclusive here). However, the results of \cm{} and \ch{} are more informative, and the effect sizes (which are almost negligible) confirm that differences between \eb{} and \eq{} are not statistically significant.\footnote{The \scmath{95\%} confidence intervals for the differences in click-through rates (\eb[\scriptsize]\smamath{-}\eq[\scriptsize]) for each company are: \cs[\scriptsize]=\scmath{(-0.363, 0.346)}; \cm[\scriptsize]=\scmath{(-0.047, 0.017)}; \ch[\scriptsize]=\scmath{(-0.003, 0.008)}.} 
\end{itemize}

\noindent
Finally, we carry out a ``two one-sided test''~\cite{schuirmann1987comparison} (or ``TOST'') to establish numerical boundaries that allow one to consider \eb{} and \eq{} to statistically have the same effectiveness. For simplicity, we carry out this test only for the ``aggregated'' results. We assume an equivalence margin of \smamath{\pm}1\%. We find that the lower bound \smamath{p}-value is \smamath{.000042}, and the upper bound \smamath{p}-value is \smamath{.0034}. Given that both of these values are below \smamath{.05}, we can conclude that there is no practically meaningful difference between \eb{} and \eq{} in leading recipients to the landing webpage.

\begin{cooltextbox}
\textsc{\textbf{Answer to RQ1.}} \eb{} and \eq{} have practically the same effectiveness at bringing a potential victim to a phishing website. Such a finding is alarming: quishing emails are harder to detect~(§\ref{sssec:quishing}) but our expectation was that they were less effective at luring users w.r.t. traditional click-through phishing emails. Our findings suggest that such an hypothesis is not true.
\end{cooltextbox}

\subsection{Qualitative Analysis of \el[\large] [RQ2]}
\label{ssec:rq2}

\noindent
There are numerous insights that can be drawn by qualitatively analysing the results pertaining to \el{}.

First, it is evident that \cs{} was the company with the highest ratio of employees that visited the landing page or submitted their credentials after reading \el{}. While the sample for \cs{} was relatively small, the effectiveness of \el{} on \cm{} seems to confirm the effectiveness of such an email to deceive employees---and are based on a much larger sample (hundreds of emails). Importantly, the impression is that \el{} tends to be much more effective than both \eb{} and \eq{} against the employees of \cs{} and \cm{} (see Table~\ref{tab:simulations_results}).

It is intriguing to observe that the percentage of ``fooled'' users for \ch{} is comparatively much lower than that of \cs{} and \cm{} (and much lower also w.r.t. \eb{} and \eq{}).\footnote{While it is possible to carry out statistical comparisons of \el[\scriptsize] w.r.t. \eb[\scriptsize] (or \eq[\scriptsize]), we refrain from doing so because there are too many differences between these emails and any test would prevent any sound conclusion. For instance, \eb[\scriptsize] and \eq[\scriptsize] were designed to impersonate the IT team urging the recipient to setup multi-factor authentication to unblock their account---which is a very important task (if true); whereas \el[\scriptsize] merely requires the employee to participate in an optional survey related to their company.} However, such a result can be due to the multi-national nature of \ch{}. Indeed, \cs{} and \cm{} are mostly based in a single country, and it is reasonable to assume that their employees may share similar views that align with the respective company's agenda. Therefore, the \el{} we crafted for \cs{} and \cm{} could have been very effective at deceiving their employees. In contrast, \ch{} is not very localized and its employees may not have a strong sense of attachment to such a company (evidence of this can be found in Tables~\ref{tab:demo_affiliation} and~\ref{tab:demo_workexp}, given that the percentage of respondents that worked for 6+ years for the same company was much higher for \cs{} and \cm{} compared to \ch{}). It is also possible that the email we crafted leveraged cues that \ch{}'s employees did not find captivating (potentially because the press releases of \ch{} may not be interesting for its employees). 
In contrast, the underlying theme of \eb{} and \eq{} (i.e., ``your account has been locked'') may have been more effective at capturing the attention of \ch{}'s employees.
Another explanation is that some employees believed the emails were legitimate but were not sufficiently motivated by the content to click, resulting in a lower click-through rate (w.r.t. \eb{} and \eq{}) despite the deception being successful at a cognitive level.

Another contributing factor may be the intrinsic difficulty of crafting a single ``generic'' phishing email (whether human- or LLM-written) that resonates across the diverse workforce of a large multinational company such as \ch{}. 
As prior work has shown~\cite{greene2018user,steves2020categorizing,sarker2024multi}, the \textit{contextual relevance} of a phishing email (e.g., in terms of timing, location, pretext, or personal relevance) plays a key role in its effectiveness. Smaller organizations are more likely to have employees who are contextually aligned~\cite{burda2023peculiar}, while large companies typically exhibit greater diversity in roles, experiences, and expectations~\cite{morris2023cultural}.

Nonetheless, it would be misleading to conclude that \el{} is ineffective for \ch{} in absolute terms. Our results demonstrate that it is possible to obtain credentials from \smamath{243} employees of a multinational company with minimal effort. The phishing email was generated using the free version of ChatGPT (in Q2 2024) and relied solely on OSINT from publicly available sources. This low-cost setup highlights the appeal of such tactics to real-world attackers.

\begin{cooltextbox}
\textsc{\textbf{Answer to RQ2.}} Using OSINT data as input to an LLM can result in phishing emails that are cheap to craft while being highly effective---especially against smaller companies.
\end{cooltextbox}

\section{PPA \& Phishing Susceptibility [RQ3]}
\label{sec:awareness}

\noindent
We first present the results of our phishing awareness questionnaire~(§\ref{ssec:awareness_results}), and then answer RQ3 via a statistical assessment (§\ref{ssec:rq3}).

\textbf{Sample description.}
Overall, we obtained \smamath{131} responses to our questionnaires (\smamath{13} for \cs{}, \smamath{82} for \cm{}, and \smamath{36} for \ch{})\footnote{The significantly lower participation rate of \ch[\scriptsize] (w.r.t. \cm[\scriptsize] and \cs[\scriptsize]) is, we believe, due to the intrinsic nature of \ch[\scriptsize]'s employees.
While participation in the phishing simulation was mandatory at \ch[\scriptsize], participation in the survey was optional, and employees who were less interested may have chosen not to participate---especially given the lack of compensation. 
In contrast, \cm[\scriptsize] and \cs[\scriptsize] are smaller companies with close-knit teams, where the corporate culture may naturally encourage participation in voluntary professional activities such as this survey.}. 
Respondents varied in age: \smamath{57} were younger than 34 years, \smamath{56} were 34--54 years old, \smamath{17} older than 55 (one preferred not to say). Our sample is also relatively well-educated, with \smamath{90} participants having a degree (BSc., MSc., or PhD). Respondents also belonged to various departments: the three most prevalent ones being IT (\smamath{36}), operations (\smamath{24}), administration (\smamath{17}). Digital devices were used at work more than 75\% of the time for \smamath{117} participants. Most of our sample (\smamath{87}) has more than ten years of work experience, with only a minority (\smamath{10}) having worked for less than two years. The complete demographic details (including the repartition across companies) are in Appendix~\ref{sapp:demographics}.

\subsection{Phishing Awareness Questionnaire: Results}
\label{ssec:awareness_results}

\noindent
Table~\ref{tab:ppa_overall} reports the results, for each company, of each factor considered in our questionnaire. 
These numbers have been obtained by averaging the responses of each question across a specific company---which we report in full in Tables~\ref{tab:ppa_attitude},~\ref{tab:ppa_behavior},~\ref{tab:ppa_training},~\ref{tab:ppa_knowledge}, in the Appendix~\ref{app:questionnaires}.

For \cs{}, the \textit{attitude towards cybersecurity} is quite high (avg=\smamath{4.223} out of 5) indicating that its employees do care about cybersecurity. However, for some specific items (see ACS3, ACS4 and ACS7 in Table~\ref{tab:ppa_attitude}), the scores are comparatively lower (\smamath{\leq}4). This could be due to \cs{} employees not being strongly confident about how to act upon arising threats. The \textit{self-reported behavior} is also somewhat high (avg=\smamath{4.0}); the lowest scored items (i.e., BHV2 and BHV3 in Table~\ref{tab:ppa_behavior}) likely stem from some uncertainty in how to integrate cybersecurity practices into their daily routines. The score for \textit{CSA training experience} is low (avg=\smamath{1.667}): this is expected because \cs{} does not carry out regular training with its employees. Moreover, for \textit{training usability}, the mediocre score (avg=\smamath{2.792}) denotes that the employees of \cs{} may not perceive training as useful. In terms of \textit{knowledge and competence gain}, there is mediocre (avg=\smamath{2.862}) with high fluctuations (variance=\smamath{1.258}): a detailed look at Table~\ref{tab:ppa_knowledge} shows that the scores of some individual items (KCG6, KCG8, KGG9) are very low (\smamath{\leq}2), due to the employees of \cs{} mistakenly deeming that some benign links were actually malicious. 
\looseness=-1

For \cm{}, the \textit{attitude towards cybersecurity} is very high (avg=\smamath{4.398}). A detailed look at the individual items (Table~\ref{tab:ppa_attitude}) reveals that \cm{}'s employees have similar difficulties as those of \cs{} (i.e., ACS3 and ACS4, both having averages \smamath{\leq}3.8). Such a tendency also pertains to \textit{self-reported behavior}: despite a higher overall score (avg=\smamath{4.168}) than \cs{}, the lowest scores pertain to the same items (i.e., BHV2 and BHV3 in Table~\ref{tab:ppa_behavior}). In contrast, for \textit{CSA training experience}, the score of \cm{} (avg=\smamath{4.091}) and of \textit{training usability} (avg=\smamath{3.982}) are substantially higher than those of \cs{}: this denotes that \cm{}'s employees believe that the training provided by their company to be useful in protecting them against emerging threats. Finally, for \textit{knowledge and competence gain}, the scores (avg=\smamath{3.884}) are generally much higher than for \cs{} indicating that the employees of \cm{} may have improved their competences after training.
\looseness=-1

For \ch{}, it stands out that the \textit{attitude towards cybersecurity} has the highest score (avg=\smamath{4.631}) among all companies; however, \ch{} also had the lowest individual scores for the same two items for which the employees of both \cm{} and \cs{} struggled (i.e., ACS3 and ACS4 in Table~\ref{tab:ppa_attitude}). The situation is similar for \textit{self-reported behavior}: \ch{} has the highest scores (avg=\smamath{4.507}) and the item with the lowest score was also the lowest for \cm{} and \cs{} (i.e., BHV3 in Table~\ref{tab:ppa_behavior}). In terms of \textit{CSA training experience} and \textit{training usability}, the scores (avg=\smamath{4.667} and \smamath{4.351}, respectively) were always above 4, denoting that the employees of \ch{} appreciate the training they receive---and do so much more than either \cs{} or \cm{}. (These results suggest the employees of \ch{} have a similar profile to those of the organization considered by Schiller et al.~\cite{schiller2024employees}). Finally, \ch{} also had the highest scores for \textit{knowledge and competence gain} (avg=\smamath{4.103}), and the two items for which the scores were the lowest (i.e., KCG3 and KCG4 in Table~\ref{tab:ppa_knowledge}) were the same ones as for \cm{}.

\vspace{1mm}

\textbox{\textbf{Summary.} The employees of all companies have a strong attitude towards cybersecurity, and their self-reported behavior shows that, in general, they are confident in their own ability to deal with cyber threats. However, while the employees of \cm{} and \ch{} appreciate the training they receive, and believe that it increases their security awareness, this is not the case for \cs{}'s employees---who believe that their training is poor and not very useful.}

\subsection{Statistical Assessment of PPA and PS [RQ3]}
\label{ssec:rq3}

\noindent
To objectively answer RQ3, we use a linear regression~\cite{montgomery2021introduction}. 
We set our hypothesis as ``The perceived phishing awareness (i.e., PPA-score) is not a statistically significant predictor of phishing susceptibility (i.e., PS-score).'' The PPA-score of each company, taken from the last row of Table~\ref{tab:ppa_overall}, is: \smamath{3.108} for \cs{}, \smamath{4.068} for \cm{}, and \smamath{4.393} for \ch{}. The PS-score of each company (computed by aggregating the ``page visited / email read'' ratios across \eb{}, \eq{}, \el{}---see Table~\ref{tab:simulations_results}) is: \smamath{38.24} for \cs{}, \smamath{15.01} for \cm{}, \smamath{7.14} for \ch{}.

After fitting a linear regression model (shown in Fig.~\ref{fig:linear} in Appendix~\ref{sapp:results}), we obtain the following results. First, the coefficient of determination is \smamath{1.0}, meaning that the model can explain possible variance in the variables (i.e., the model is a perfect fit). Second, the slope is \smamath{-24.2}, with an intercept of \smamath{113.45}. Third, and more importantly, the \smamath{p}-value is \smamath{<.001}. Therefore, we must reject our hypothesis. We further validated this finding via a Spearman's Rank Correlation test~\cite{gauthier2001detecting}, obtaining  \smamath{\rho}=\smamath{-1.0} with \smamath{p}-value=\smamath{0}. 

Given that the PPA score ranges between 1--5, we use our linear regression model to estimate the corresponding PS-score. For instance, a hypothetical company with a PPA-score=\smamath{1}, an extremely low level of perceived phishing awareness, would be expected to have a PS-score=\smamath{89.26}, indicating that nearly 90\% of its employees would click through to the phishing webpage upon receiving an email such as \eb{}, \eq{}, or \el{}.
In contrast, any company with a PPA-score\smamath{\geq}4.7 would have a PS-score\smamath{\leq}0 (i.e., nobody would land on the phishing webpage if they received an email similar to our \eb{}, \eq{}, or \el{}). Of course, these are just extreme scenarios.

\vspace{-1mm}

\begin{cooltextbox}
\textsc{\textbf{Answer to RQ3.}} We found a strong correlation between the perceived phishing awareness (PPA) and the phishing susceptibility of a company's employees. By estimating the PPA of (a subset of) the employees of a given company, it is possible to predict their phishing susceptibility---potentially of the entire company. 
\end{cooltextbox}

\begin{table}[!t]
    \centering
    \caption{\textbf{Summary of the PPA (mean and variance) for each company.} \textmd{We provide the data used to derive these results in Appendix~\ref{app:questionnaires}.}}
    \vspace{-2mm}
    \resizebox{0.99\columnwidth}{!}{
        \begin{tabular}{l?c|c||c|c||c|c}
            \toprule
            & \multicolumn{2}{c||}{\cs{}} & \multicolumn{2}{c||}{\cm{}} & \multicolumn{2}{c}{\ch{}}\\
            \cline{2-7}
             & Mean & Var. & Mean & Var. & Mean & Var. \\
            \midrule
            
            Attitude towards Cybersecurity & $4.223$ & $0.596$ & $4.398$ & $0.416$ & $4.631$ & $0.316$ \\
            Self-reported Behavior & $4.000$ & $0.728$ & $4.168$ & $0.569 $ & $4.507$ & $0.387$ \\
            CSA Training experience & $1.667$ & $0.222$ & $4.091$ & $0.550$ & $4.667$ & $0.333$ \\ 
            Training Usability & $2.792$ & $1.972$ & $3.982$ & $0.797$ & $4.351$ & $0.638$ \\
            Knowledge \& Competence Gain & $2.862$ & $1.258$ & $3.884$ & $1.197$ & $4.103$ & $1.176$ \\
            \rowcolor{gray!30} OVERALL (PPA-score) & $3.108$ & $0.955$ & $4.068$ & $0.788$ & $4.393$ & $0.671$\\

            \bottomrule
        \end{tabular}
    }
    \label{tab:ppa_overall}
    \vspace{-4mm}
    
\end{table}
\section{Discussion and Critical Analysis}
\label{sec:discussion}

\noindent
We draw lessons learned from our research~(§\ref{ssec:lessons}), discuss limitations of our study~(§\ref{ssec:limitations}), and suggest avenues for future work~(§\ref{ssec:future}). In doing so, we also position our findings within related work. 

Moreover, we further examine our findings (discussing, e.g., the report-rate and credential-submitted) in Appendix~\ref{app:considerations}.

\subsection{Major Findings and Lessons Learned}
\label{ssec:lessons}

\noindent
We distill three lessons learned that can be used as a foundation for future research on the threat of phishing in organizations.

The first lesson learned is that \textit{quishing emails are dangerous}. The findings of RQ1 show that quishing emails have the same effectiveness as traditional click-through emails in ``hooking'' users to a phishing website. Such a result, combined with the intrinsic and subtle characteristics of quishing emails (§\ref{sssec:quishing}), makes this form of phishing attack particularly problematic. In a sense, quishing emails are more threatening than other types of phishing emails---such as those leveraging click-through buttons or plaintext URLs, both of which can be detected more easily via automated mechanisms~\cite{subramani2022phishinpatterns,petelka2019put}. Some works (e.g.,~\cite{vaithilingam2024enhancing, yong2019survey}) propose ways to detect quishing emails; yet, as we argued (in §\ref{sssec:quishing}), and as also confirmed by the increasing popularity of QR-codes as a phishing-email vector~\cite{talos2024qr}, currently deployed defenses do not seem to be effective. We endorse future studies to put more attention to quishing.

The second lesson learned is that \textit{LLMs, if fed with OSINT, can be very effective at crafting phishing emails}. For instance, for \cm{}, \smamath{31.5\%} (resp. \smamath{14.9\%}) of the employees who read the email visited the landing webpage (resp. submitted their credentials). This result is worrying, given that \el{} was created without exploiting tactics such as urgency or sense of loss. Notably, for \cm{}, this email was up to 5 times more effective than either \eb{} and \eq{} at ``persuading'' users (likely because \cm{}'s employees already use two-factor authentication and do not need to set it up).  Our results can be compared to those of some (unpublished) works that measured the effectiveness of OSINT-fed LLM on employees of a single organization: in both ~\cite{heiding2024evaluating,ibm2023llm}, \smamath{\approx}\smamath{11\%} of the employees who opened the email visited the landing webpage; these numbers are higher than those we obtained for \ch{}, but much lower than those of \cm{} and \ch{}. Such differences call for more work to explore the effectiveness of OSINT-fed LLMs to write phishing emails against employees of a wide range of companies.
Nonetheless, given that {\small \textit{(i)}}~our approach to craft \el{} can be applied to most large companies, and that {\small \textit{(ii)}}~LLMs are getting increasingly better at elaborating and generating data~\cite{schoenegger2024wisdom}, we can expect that feeding OSINT to LLMs will become commonplace in crafting phishing emails, which can also be targeting the specific employee rather than their company (as done in~\cite{heiding2024evaluating}). We thus endorse future work to develop ways to counter LLM-written phishing emails. 

The third lesson learned is that the \textit{employees' perceived phishing awareness can be a predictor of a company phishing susceptibility}. Importantly, our findings seem to hold across different companies. The outcome of RQ3 complements that of recent works~\cite{schiller2024employees,lain2024content}: phishing training may provide a false sense of security because employees who did well during training still fell for some phishing traps. Our study, however, had a different purpose, given that our PPA-score is computed by accounting for five different factors---which are a superset of the ``knowledge and competence gain'' from CSA training~\cite{chaudhary2022developing}. Nonetheless, our fine-grained results (in Appendix~\ref{sapp:results}) enable future work to consider different elements to derive a different PPA-score.
We stress that the findings of RQ3 were possible thanks to our collaboration with companies which allowed us to {\small \textit{(i)}}~conduct phishing simulations to gauge the phishing susceptibility, and {\small \textit{(ii)}}~disseminate a survey to measure the perceived phishing awareness of their employees. Accomplishing such an effort in three different contexts required us to overcome many challenges (see, e.g., Appendix~\ref{app:technical}), which may explain why we could not find papers that shared a similar goal. For instance, some very recent works (e.g.,~\cite{ho2025understanding,lain2024content,schiller2024employees}) only consider a single company. 

\textbox{\textbf{Remark.} Our findings depend {\small \textit{(a)}}~on our chosen companies---predominantly based in Europe, and pertaining to three specific businesses (see Table~\ref{tab:companies}); and {\small \textit{(b)}}~on our emails---which could have been created in a plethora of other ways. For instance, the generic idea for \eb{} and \eq{} was to imitate an email related to institutional credentials (similarly to, e.g.,~\cite{yeoh2022simulated,lain2022phishing}), but there exist other hooks that could have been used to compare a click-through button with a QR code (e.g.,~\cite{schiller2024employees,sarno2022so}). At the same time, there are many ways (e.g.,~\cite{roy2024chatbots,langford2023phishing,heiding2024evaluating}) in which LLMs can be used to craft the email valid for RQ (ours is inspired by~\cite{ibm2023llm}). Therefore, \textit{we do not seek to generalize our conclusions}, nor claim generalizability.}.

\subsection{Threat to Validity and Limitations}
\label{ssec:limitations}

\noindent
Our study is complex and entailed carrying out a number of experiments in different companies. Let us discuss the most evident issues that could have threatened the validity of our conclusions.

First, \textit{the fact that, for \eq{} in \ch{}, we did not have complete control of the experiment}. Since \ch{} had recently carried out a quishing simulation on their own, we could not carry out another experiment so soon. Moreover, and unfortunately, \ch{} did not collect data pertaining to the ``submitted credentials'' for their own quishing simulation. However, neither of these facts threaten our conclusions: 
\begin{itemize}[leftmargin=*]
    \item For RQ1, we compare the effectiveness of a quishing vs.\ click-through phishing email in leading a user to a malicious webpage. Therefore, what the user does \textit{after} landing on such a webpage is outside the scope of RQ1 (see also §\ref{ssec:approach}). Hence, lack of data on the submitted user credentials is irrelevant for RQ1.
    \item For RQ3, we consider the PS-score, i.e., the ratio between the ``visited webpage / email read'', which does not depend on the number of users that submitted their credentials.
\end{itemize}
Finally, for RQ2, we carry out a qualitative analysis which is based on a different email (i.e., \el{}), for which we have all data.

Second, \textit{the fact that the timeline of our analysis may have affected our results.} Indeed, we first carried out the phishing simulations, and a couple of days later we carried out the survey for assessing the PPA (see §\ref{sec:method}). Hence, it is possible that some answers to the PPA-related questions had been influenced by the recent phishing simulation. However, such a possibility only pertains \cs{}: the two other companies are used to carry out phishing assessments quite often. Nevertheless, such a possibility does not impact our conclusions whatsoever for RQ1 and RQ2, and it may have had only a minor effect on RQ3 (because \cs{} is the smallest company).

Third, \textit{the fact that we have no information on ``who'' participated in the PPA survey}. Therefore, we do not know how such participants performed in the previous phishing simulations. If we could know what the participants of our PPA survey did with the respective emails, we could carry out a more fine-grained assessment. Unfortunately, we do not have access to such (confidential) data. Nonetheless, we are unsure of how such data could be fairly obtained in the first place: having an employee perform the PPA right after the phishing simulation may bias the results; moreover, we could not force an employee to participate in the PPA survey (which is voluntary), hence complete coverage may be impossible to attain. 

Fourth, and extending the previous point, \textit{the small sample size for RQ3}. Only \smamath{131} employees participated in our PPA survey, whereas we sent \smamath{71\,309} emails for our phishing simulations. When answering RQ3, we are using the relatively small sample size (i.e., a few dozen employees per company) as a representative indicator of the entire company---which is a gross generalization. Moreover, only three datapoints have been used to answer RQ3 (i.e., we have the PS-score and PPA-score for three companies) for our linear regression model. However, to validate our results, we also attempted to consider all companies as a single entity (with a PS-score of \smamath{7.3\%} and a PPA-score of \smamath{4.06}): this yields a fourth datapoint that can be used to compute the linear regression anew---which results in a \smamath{p}-value=\smamath{.036}\smamath{<}\smamath{.05} which still supports our conclusion. Nonetheless, it is known (see, e.g.,~\cite{lain2024content,schiller2024employees}) that finding volunteers for similar studies is tough.

\subsection{Future Work}
\label{ssec:future}

\noindent
Our study opens new grounds for future research addressing the problem of phishing in organizations. In what follows, we emphasize three areas that deserve particular attention.

\textbf{Countermeasures.} 
Our findings (§\ref{ssec:lessons}) highlight the need for dedicated mitigations. 
First, defenses against quishing emails are essential although difficult to implement~\cite{ford2024feasibility}. 
Server-side approaches are resource-intensive, while client-side solutions offer a promising alternative (e.g., QR-code scanners that automatically flag suspicious URLs~\cite{rafsanjani2023qsecr}). 
Second, addressing malicious content generated by OSINT-fed LLMs remains an open challenge. Humans struggle to distinguish between human- and LLM-written text~\cite{frank2024representative}. In the phishing context, one option is to apply detectors of machine-generated content (e.g.,~\cite{hans2024spotting}), though they have known limitations. We expand on these and other potential defenses in Appendix~\ref{app:considerations}. 
We encourage future research to build on our findings, both as motivation and as a foundation for developing effective countermeasures.
\looseness=-1

\textbf{More LLMs.} 
Our study relied on GPT-3.5 Turbo, which was the free version of ChatGPT available at the time (early 2024). However, the LLM landscape is rapidly evolving, and newer models often surpass their predecessors. This ongoing progress also benefits attackers, who gain access to increasingly capable (and often free) tools against which countermeasures remain limited. We anticipate that LLMs’ ability to craft persuasive, and thus more deceptive, phishing emails will only improve over time~\cite{salvi2024conversational}, thus making findings reported in this paper a `lower bound' of what is to come.
Future research should evaluate the phishing potential of emerging models, including newer versions of ChatGPT and offerings from other vendors (e.g., Claude, Llama, Gemini). Understanding the capabilities of these models, how they fit into offensive practices, and how they affect attackers' \textit{modus operandi} and capabilities is essential for designing effective defenses.
LLM providers have begun to acknowledge these risks (see~\cite{roy2024chatbots}), and---ideally---will support efforts to study and mitigate such threats.

\textbf{More Organizations.} 
Our study focuses on three companies operating in the financial, hospitality, and manufactoring sectors. However, \textit{our considered phishing threats can affect virtually any organization}---as evidenced by ProofPoint's recent report~\cite{proofpoint2024phish}. For instance, prior work has shown the simplicity of carrying out OSINT operations against employees of critical infrastructures~\cite{edwards2017panning}, which can be leveraged to craft specific phishing emails against similar organizations. More generally, the public sector (including, e.g., higher education~\cite{jensen2017training}, healthcare~\cite{priestman2019phishing,gordon2019assessment}, or government~\cite{koddebusch2022exposing}) is at constant risk of phishing attacks. Whereas our findings do not directly map to these organizations (since they have a different workforce), our research methods can be applied to study these complementary contexts---an intriguing avenue for future work.
\section{Conclusions and Recommendations}
\label{sec:conclusion}
\noindent
Our study is a stepping stone towards understanding the impact of quishing and LLM-based phishing emails across organizations.

We found that embedding malicious QR-codes in phishing emails has the same effectiveness at luring users to a landing webpage as a traditional click-through button. This result is alarming, given that quishing emails can bypass most filters (as also demonstrated by our experiment). We recommend security developers and researchers alike to prioritize the implementation of automated defenses that can mitigate the widespread usage of malicious QR codes. We also promote the inclusion of quishing in CSA training exercises.

The results of our assessment of LLM-based phishing emails, as well as those of our PPA survey, should also serve as an inspiration for future work. Ultimately, and unfortunately, there are many ways to use LLMs to craft phishing emails. Moreover, every company has different employees. By providing all our results and methods, we hence enable downstream research to carry out similar assessments and compare their results with ours---thereby further expanding our understanding of emerging phishing threats.

\section*{Acknowledgements}
We thank the anonymous reviewers and our considered companies. This work has been partially supported by the Hilti Foundation, and the INTERSECT project, Grant No. NWA.1162.18.301, and the SeReNity project, grant no. CS.010, funded by the Netherlands Organization for Scientific Research (NWO). Any opinions, findings, conclusions, or recommendations expressed in this work are those of the author(s) and do not necessarily reflect the views of NWO.

\bibliographystyle{ACM-Reference-Format}


\appendices
\section{Technical details (and challenges)}
\label{app:technical}
\begin{figure}[!t]
    \centering
    \includegraphics[width=0.85\columnwidth]{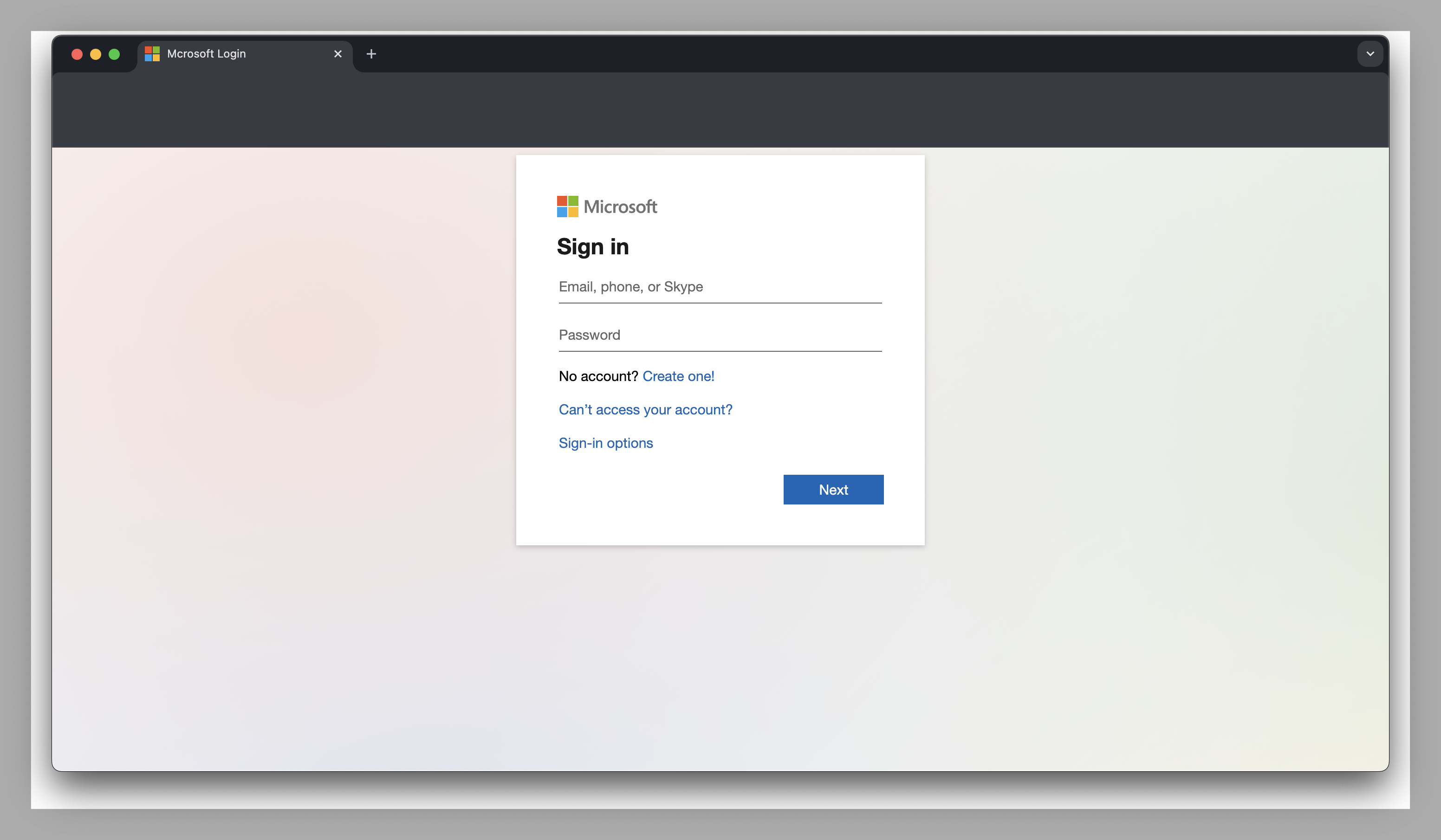}
    \vspace{-2mm}
    \caption{\textbf{Landing page.} \textmd{All of our emails would point to a webpage with a similar design as this one, showing the typical ``Microsoft login''.}}
    \label{fig:landing}
    \vspace{-3mm}
\end{figure}

\noindent
We expand the information provided in §\ref{ssec:setup}, which covers our setup for \cs{} (for which we show our custom landing webpage in Fig.~\ref{fig:landing}). Additionally, we provide the prompts used to craft \el{} in Table~\ref{tab:prompts}.

\begin{table}[t!]
    \centering
    \caption{Sequence of Prompts used to craft \el[\scriptsize]. \textmd{Text in regular font are not part of the prompt; the last prompt is optional. We do not show the prompts used to ``jailbreak'' the model (to avoid helping attackers).}}
    \label{tab:prompts}
    \vspace{-3mm}
    \setlength{\tabcolsep}{5pt} 
    \renewcommand{\arraystretch}{1.2}
    \begin{tabular}{|c|>{\setstretch{0.4}\raggedright\arraybackslash}p{7.4cm}|}
        \hline
        \textbf{\#} & \textbf{Prompt} \\ \hline
        1 & \prompt[\scriptsize]{Please help me summarize the weaknesses this company has according to this employer rating website.} {\scriptsize [Extra input: data extracted from Kununu]} \\ \hline
        2 & \prompt[\scriptsize]{If I were an attacker, which weakness would be the best to leverage in a phishing attack?} \\ \hline
        3 & \prompt[\scriptsize]{Please give me one concrete example of a potential phishing mail leveraging this weakness.} \\ \hline
        4 & \prompt[\scriptsize]{Please analyse these postings for me and give me the 5 most common topics that this company cares about.} {\scriptsize [Extra input: data extracted from LinkedIn]} \\ \hline
        5 & \prompt[\scriptsize]{Please write me a brief introduction to a company survey directed at employees regarding the latest company efforts in relation to [topic from prompt \#4] at [company]. The introduction is meant to accompany the link to the survey. Here is some additional information the employees are already aware of.} {\scriptsize [Extra input: text from press releases]} \\ \hline
         & \prompt[\scriptsize]{Shorter please} {\scriptsize [Note: only added if the output was longer than 100 words so that it would still be readable]} \\ \hline
    \end{tabular}
    \vspace{-4mm}
\end{table}

\subsection{Microsoft Defender}
\label{sapp:defender}

\noindent
The simulations for \cm{} and \ch{} leveraged the ``Microsoft Defender Attack Simulation Training'' (MADST) module, which is part of Microsoft’s Office 365 licensing for large enterprises~\cite{defender}.

\begin{figure}[!htbp]
    \centering
    \includegraphics[width=0.9\columnwidth]{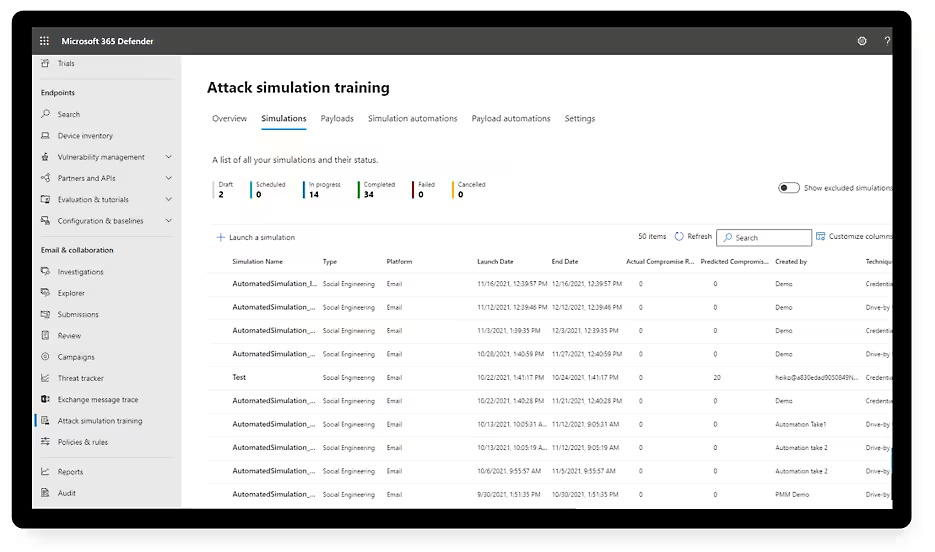}
    \vspace{-3mm}
    \caption{\textbf{Interface of Microsoft Defender Attack Simulation module.} \textmd{This is just an example, no confidential information is shown.}}
    \label{fig:msd}
    \vspace{-3mm}
\end{figure}

This module helps organizations to run realistic simulations in their workplace using Microsoft’s own ecosystem. We show in Fig.~\ref{fig:msd} a sample visualization of its interface. For \cm{} and \ch{}, we used the same module deployed by the respective security team. Given the highly-confidential nature of our research, we spent a lot of time discussing with \cm{} and \ch{} so that we could find an agreement on how to use their framework for our experiments. 

Among the greatest challenges we encountered was integrating the QR-code email simulation in MADST. Indeed, MADST does not support QR-code emails natively. Hence, we had to deploy a dedicated webapp that would create a custom QR code from the specific link of an email; doing this was not simple from a bureaucratic viewpoint, given the ``external'' nature of such a webapp. 

Nonetheless, the MADST module supports various ``attack scenarios''. For our experiments, we opted for the ``credential harvesting scenario'' (see~\cite{credentialharvest}), since it aligned with our goals and was also supported by GoPhish. 

Finally, for \ch{} and \cm{}, we also relied on their ``feedback'' page (similar to the one shown in~\cite{feedback}: we cannot show the actual one due to NDA) that informs users who submitted their credentials that they have been ``phished''. We did not do this for for \cs{} since it was not deemed necessary by their representatives. Indeed, due to the small size of \cs{}, it was possible to directly reach out to each user who submitted their credentials and let them know that they ``fell'' in a phishing trap. Regardless, such a discrepancy has no impact on the goal of our study.

\subsection{GoPhish}
\label{sapp:gophish}

\begin{figure}[!htbp]
    \centering
    \includegraphics[width=0.9\columnwidth]{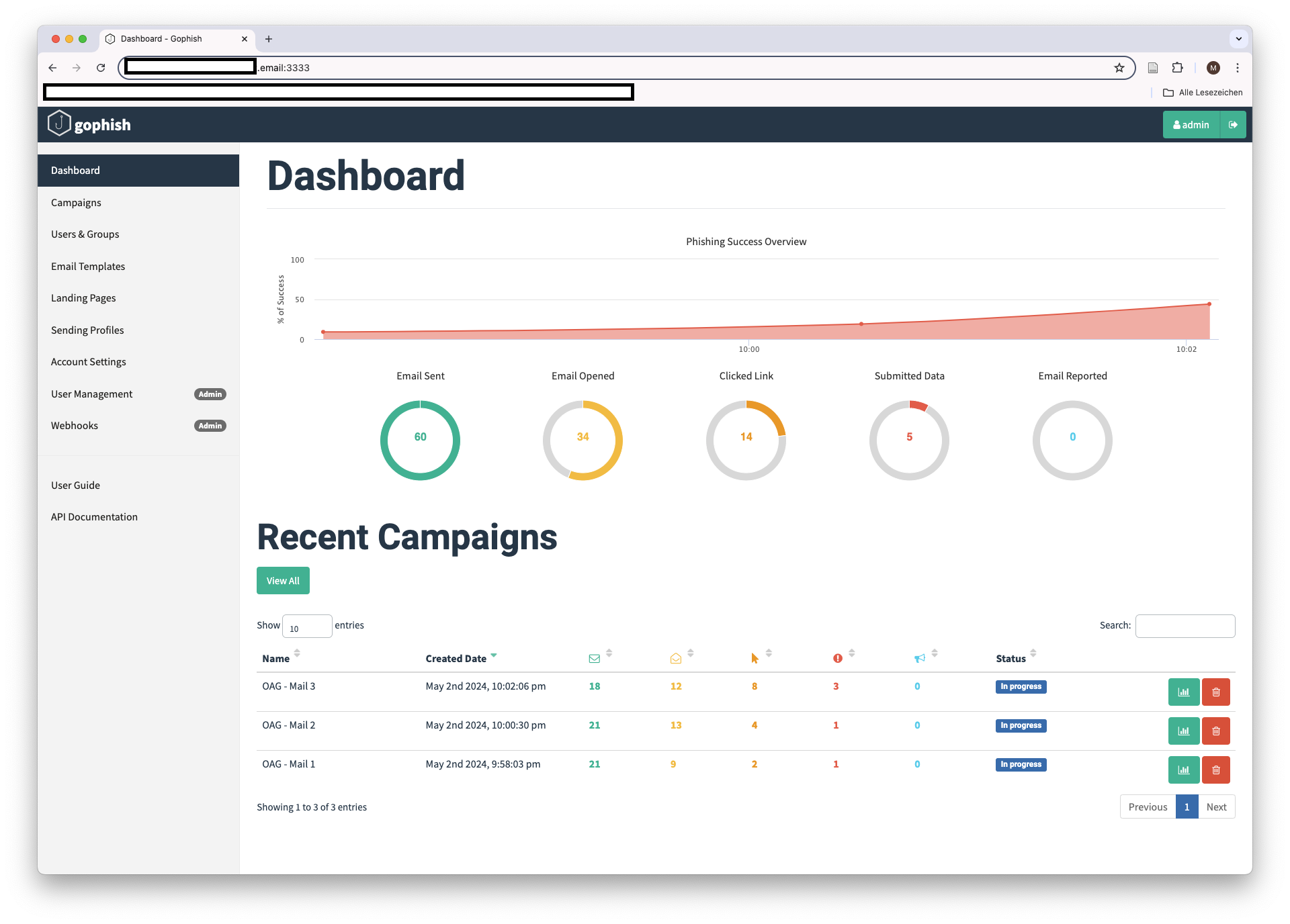}
    \vspace{-3mm}
    \caption{\textbf{Interface of GoPhish.} \textmd{This is just an example, no confidential information is shown.}}
    \label{fig:gophish}
    \vspace{-2mm}
\end{figure}

\noindent
GoPhish is an open-source phishing framework written in the programming language ``Go''~\cite{gophish}. GoPhish seeks to make phishing
assessments and training available and accessible for everyone. We provide in Fig.~\ref{fig:gophish} a sample of GoPhish dashboard. We used GoPhish for \cs{}: unfortunately, setting up GoPhish for our experiment revealed to be much more complex than what we had foreseen.

\textbf{Internet deployment.} We deployed our instance of GoPhish on a virtual private server (VPS) ``exposed'' to the internet, since we needed it to be operational and accessible during the entire time of our simulations. To this end, we licensed a small (specs: OS=Ubuntu 23.10; CPU Type=Regular Intel (1 CPU); RAM=512MB; SSD=10GB; Location=Germany) VPS with a well-known cloud-service provider, which cost us 4\$ per month. We also purchased a domain (which would resemble \cs{}'s name, to which we added ``.email'' to it) with a well-known domain registrar, which cost us 9\$. Such a cost was necessary to increase the realistic fidelity of our campaign: otherwise, users could be suspicious if, e.g., they saw IP addresses or weird domains in the emails they received. To further increase the credibility of our infrastructure, we also set up an SSL certificate for our domain (we used ZeroSSL, which provided a free service for the first 90 days).

\textbf{SMTP relay.} The provider of our VPS automatically blocks the SMTP protocol (``to avoid misuse by criminals''). Unfortunately, such a protocol was, of course, required for our simulation. To overcome this challenge, we set up an SMTP relay~\cite{smtprelay}. This required us to authenticate our domain by creating four CNAME records, and add them to the VPS. This way, we verified our domain and were able to set up the sender identity (which we chose as described in §\ref{sssec:common}). The SMTP relay we chose was free for up to 100 emails per day, so we did not incur in any costs---given the small size of \cs{}.

\textbf{Blocked emails.} Once we set up the abovementioned infrastructure, we began doing some tests by sending some emails. Unfortunately, we found that such emails, when sent to Gmail addresses, were blocked by Google's spam filters and put in the ``Junk'' folder; other email providers blocked the emails entirely (no email was received even after 48 hours). To overcome this problem, we agreed to have \cs{} enter the sender of our emails among the whitelisted senders for \cs{} mail client. However, the issues did not stop here: when we tested the landing page we created for \cs{} (hosted on the VPS), we found out that it had been blocked by Google's SafeBrowsing. We reached out to Google, reporting the page as benign: thankfully, the block was lifted after 24hours.

\begin{figure}[!htbp]
    \centering
    \includegraphics[width=0.9\columnwidth]{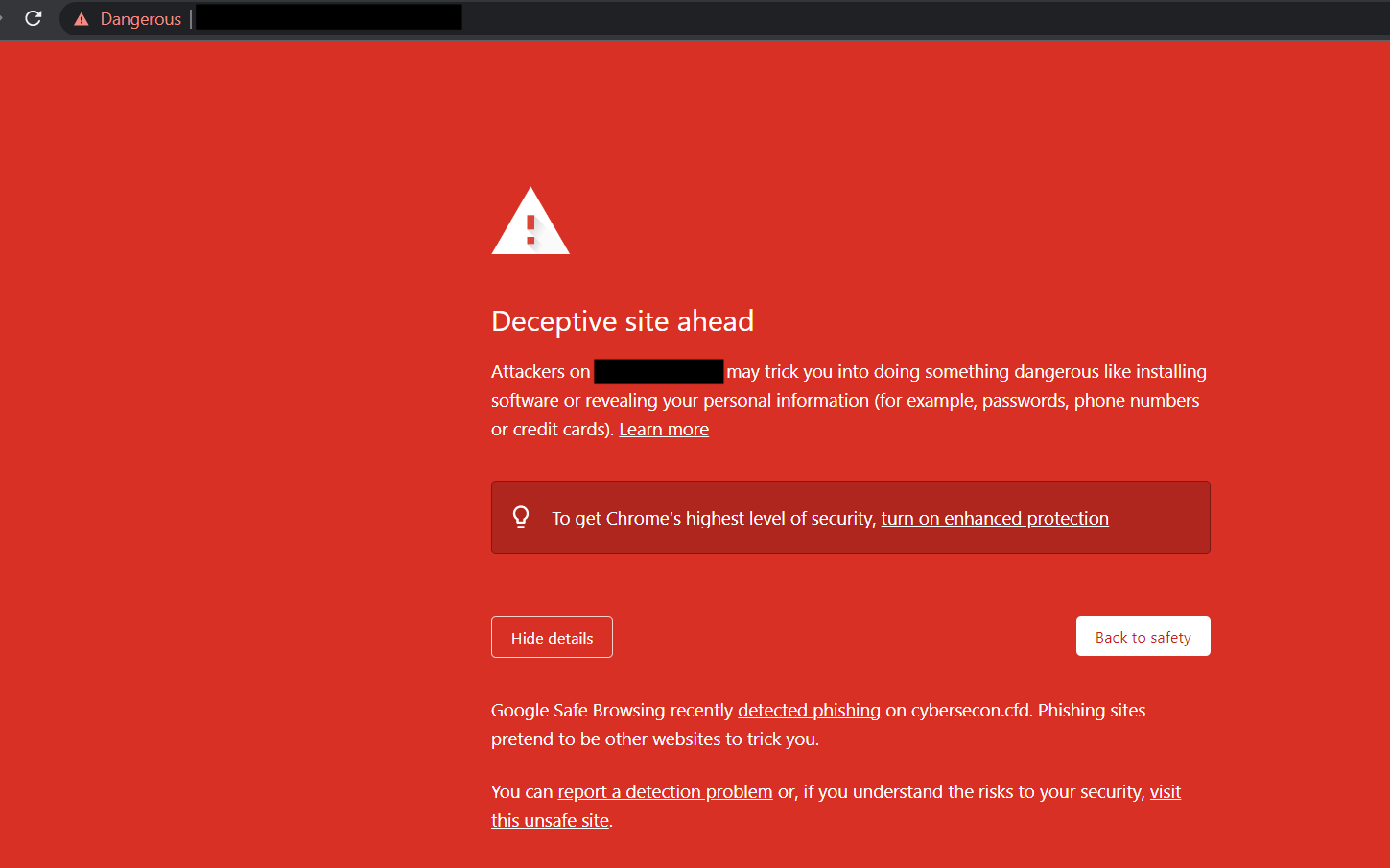}
    \vspace{-3mm}
    \caption{\textbf{Our landing page was initially blocked by Google SafeBrowsing.} \textmd{We reached out to Google who lifted the block after 24 hours.}}
    \label{fig:block}
    \vspace{-3mm}
\end{figure}

\begin{figure*}[!htbp]
\centering

    \begin{subfigure}[!htbp]{1\columnwidth}
        \centering
    \frame{\includegraphics[height=4.5cm]{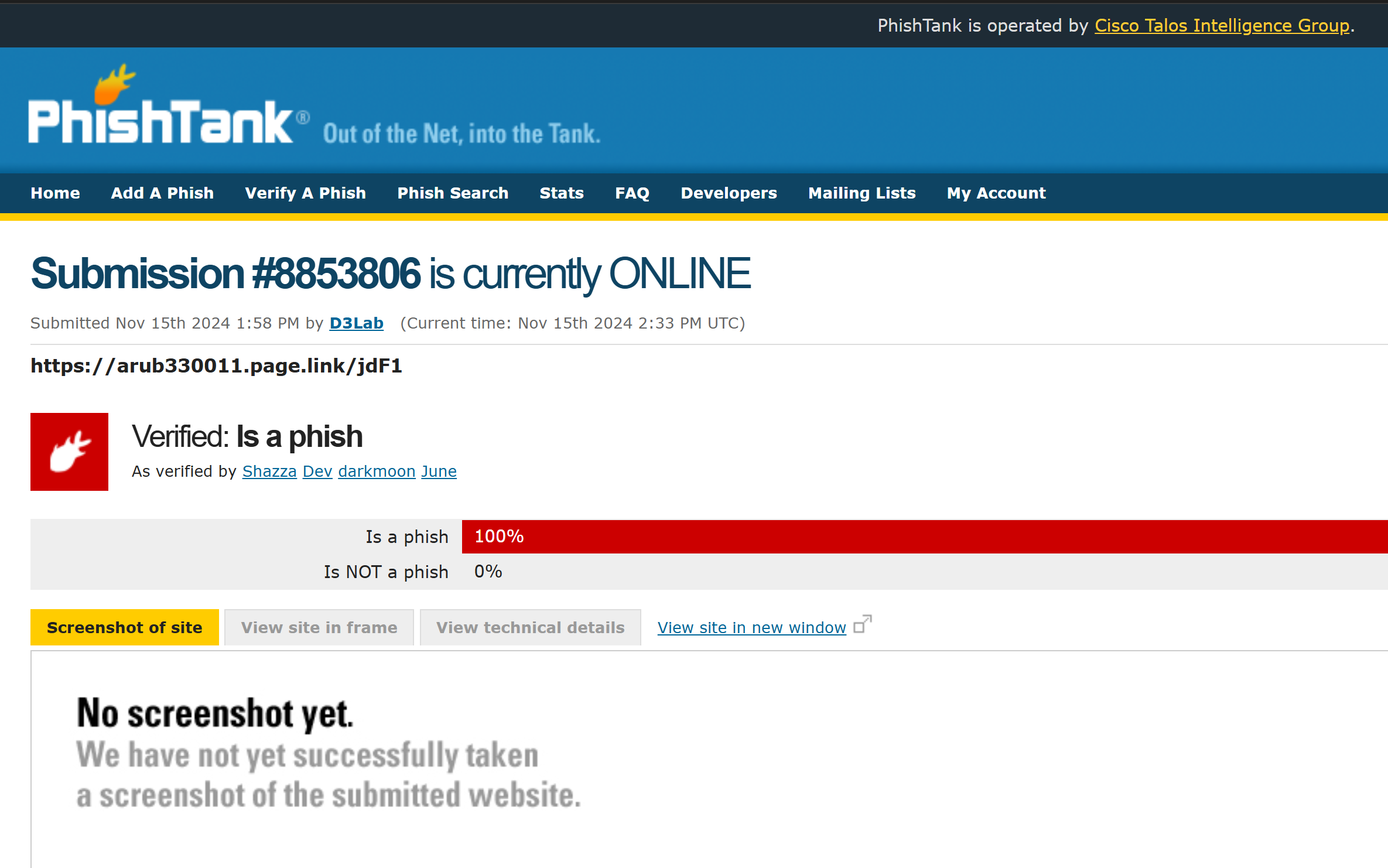}}
    \caption{\textmd{Details of the malicious URL (https://arub330011.page.link/jdF1) according to Phishtank~\cite{phishtank} (in November 2024).}}
    \label{sfig:phishtank}
    \end{subfigure} 
    \begin{subfigure}[!htbp]{0.3\columnwidth}
        \centering
    \frame{\includegraphics[height=2.5cm]{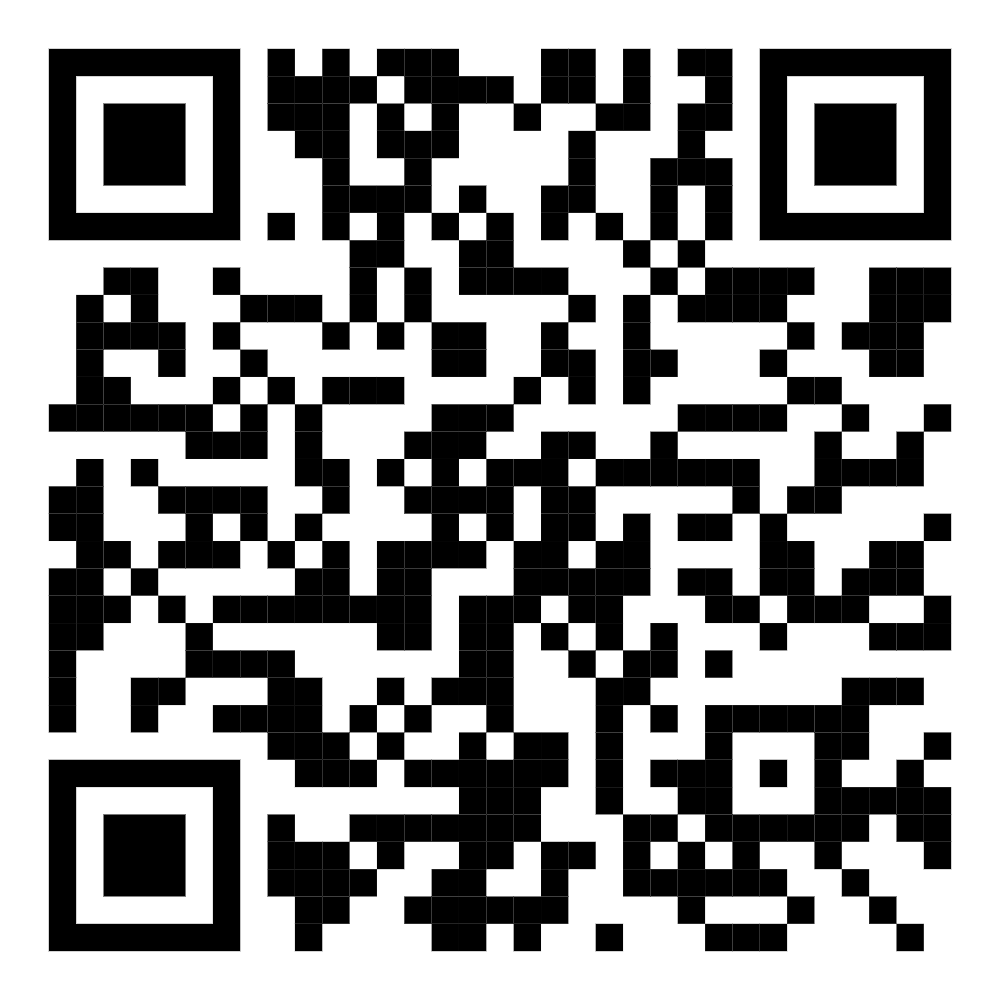}}
    \caption{\textmd{QR-code of the malicious URL used as a basis for this experiment.}}
    \label{sfig:qr}
    \end{subfigure} 
    \begin{subfigure}[!htbp]{0.7\columnwidth}
        \centering
    \frame{\includegraphics[height=4.5cm]{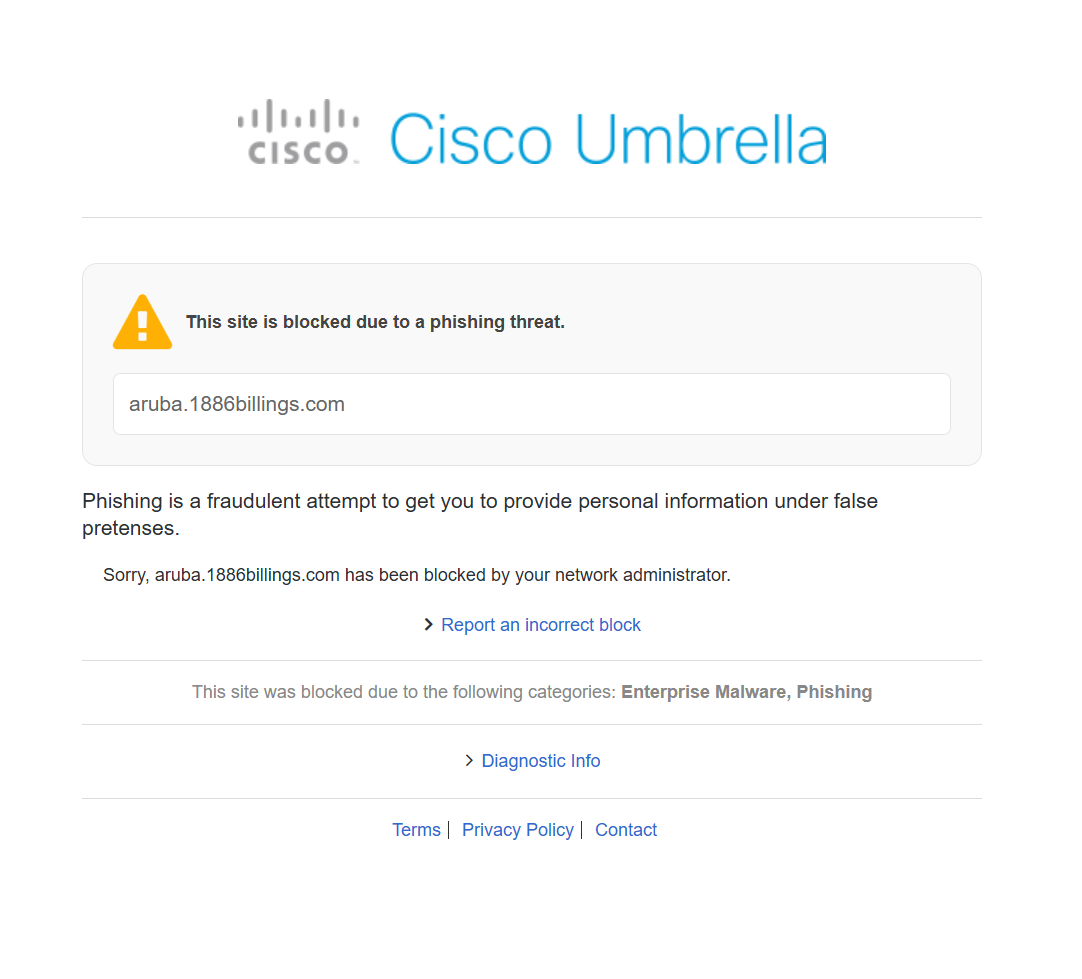}}
    \caption{\textmd{Verification that the URL was known to be malicious by well-known providers (e.g., CISCO).}}
    \label{sfig:blocked}
    \end{subfigure}

    \vspace{-2mm}
    \caption{\textbf{Original QR-code test: preliminaries.} \textmd{We took a URL pointing to a phishing webpage from Phishtank (Fig.~\ref{sfig:phishtank}), we generated the corresponding QR code (Fig.~\ref{sfig:qr}) and also checked that the webpage had been included in operational blocklists (Fig.~\ref{sfig:blocked}) used by popular browsers.}}
    \label{fig:source}
    \vspace{-3mm}
\end{figure*}

\section{Do quishing emails evade operational detectors?}
\label{app:test}

\begin{figure}[!htbp]
\centering

    \begin{subfigure}[!htbp]{0.99\columnwidth}
        \centering
    \frame{\includegraphics[height=2.5cm]{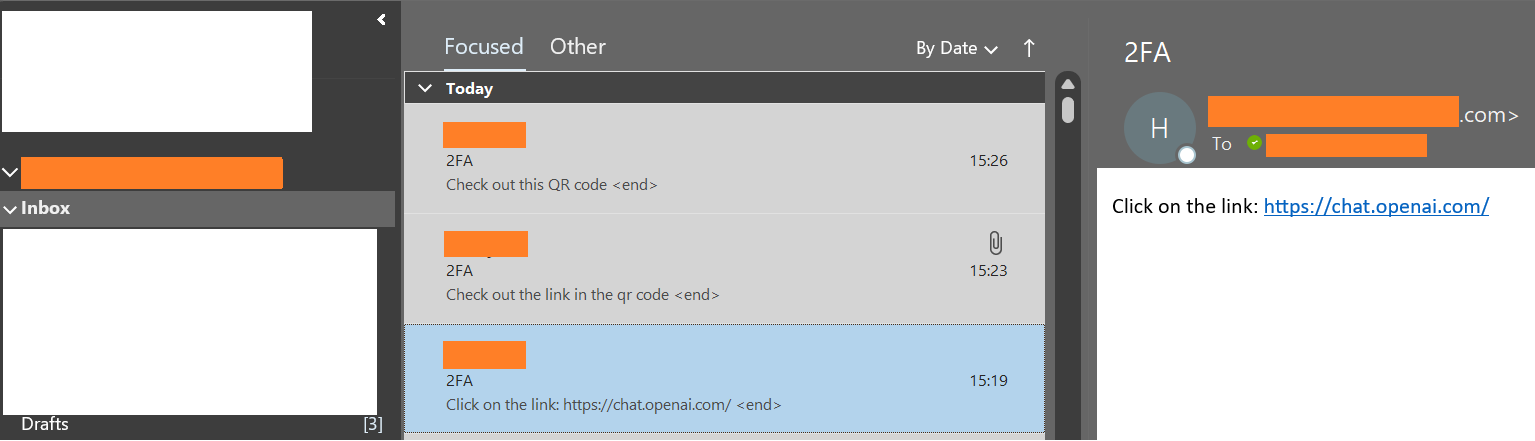}}
    \caption{\textmd{Aside from the ``benign'' email (sent at 15:19), the two ``quishing'' emails (one having the QR code as attachment, the other embedded in the email's content) were not put in the junk folder---thereby evading the phishing/spam detection filter.}}
    \label{sfig:inbox}
    \vspace{1mm}
    \end{subfigure} 
    \begin{subfigure}[!htbp]{0.99\columnwidth}
        \centering
    \frame{\includegraphics[height=1.7cm]{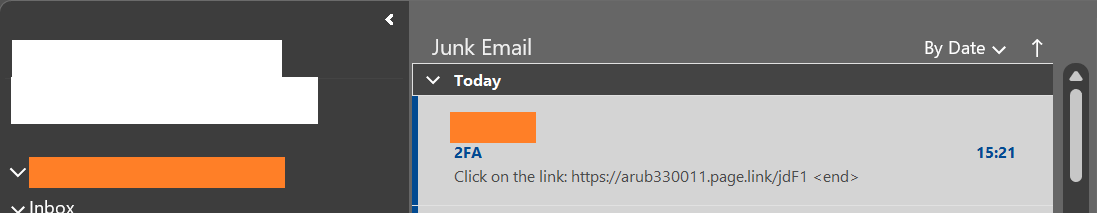}}
    \caption{\textmd{The only mail put in the ``junk'' folder was the one with the URL in plaintext.}}
    \label{sfig:junk}
    \end{subfigure}

    \vspace{-2mm}
    \caption{\textbf{Original QR-code test: results.} \textmd{We sent four emails to the institutional email address of one of the authors, managed by Microsoft (i.e., the same provider of the companies considered in our paper---see §\ref{ssec:companies}).}}
    \label{fig:client}
    \vspace{-5mm}
\end{figure}

\noindent
To provide real-world evidence that QR-code emails are ``more stealthy'' than traditional URL-based phishing emails, we carried out an original experiment on an \textit{operational detector}.\footnote{The infrastructure that supports the institutional email services of (some of) the authors of this paper is provided by Microsoft, which integrates anti-phishing tools~\cite{antiphishing}.}
In November 2024, we retrieved a malicious URL (i.e., {\small https://arub330011.page.link/jdF1}) from phishtank~\cite{phishtank} (see Fig.~\ref{sfig:phishtank}). We first verified that it was included among common blocklist: we tried visiting the URL, and we were shown a warning webpage (see Fig.~\ref{sfig:blocked}). Then, we generated a QR-code for such a URL (see Fig.~\ref{sfig:qr}). At this point, we sent four emails---all from the same email account (i.e., the personal Gmail account of one of the authors) to the same email account (i.e., the institutional email account of the same author). Specifically:
\begin{itemize}[leftmargin=*]
    \item The first email was just a sanity check, and it simply included a link to a well-known (benign) website, asking the recipient to ``click on the link''. This email, as expected, was put in the \textit{inbox folder} of the recipient account.
    \item The second email was the URL-based phishing email: it was the same as the first email, but instead of the benign link we put the malicious link mentioned above. This email was put in the \textit{junk folder} of the recipient account.
    \item The third email was an exemplary quishing email in which the QR code was provided as an image attachment; the text invites the reader to ``check out the link in the qr code''. This email was put in the \textit{inbox folder} of the recipient account.
    \item The fourth email was also a quishing email in which we put the QR code in the body of the email (i.e., as an HTML object). This email was put in the \textit{inbox folder} of the recipient account.
\end{itemize}
The four emails above all had the same subject (``2FA'') and had been sent within a timespan of 7 minutes. It is possible to visualize the results of this experiment in Fig.~\ref{fig:client}. 
These results demonstrate that both quishing emails have ``evaded'' a commercial phishing/spam filter---despite the corresponding URL-based phishing email being (correctly) deemed as junk.

\section{Systematic Literature Review}
\label{app:systematic}
\noindent
Quishing has been somehow overlooked by prior research---at least from the viewpoint of papers focused on user studies. 

Indeed, we have systematically analysed the 2014--2024 proceedings of 11 top-venues related to Security, Human Factors and the Web: WWW, S\&P, EuroS\&P, CCS, USENIX SEC, NDSS, AsiaCCS, ACSAC, IMC, WSDM, CHI. We searched for full papers (excluding, e.g., workshops) having ``phish'' in the title and found 66 papers. Then, we inspected their text, searching for occurrences of the term ``qr''. We found only two papers with such a string: \cite{pourmohamad2024deep} (where ``qr'' was mentioned only once) and~\cite{roy2024chatbots} (here, it occurs 13 times). Neither of these, however, carried out user studies. 

More generally, however, we can state that QR-code phishing is not a commonly researched theme among these 11 top-tier venues.
\section{Additional Considerations}
\label{app:considerations}
\noindent
We provide some additional critical remarks on our research, and further support our arguments with original analyses.

\textbf{Credentials submitted.}
For RQ1, we did not consider what happens after the user lands on the webpage. A look at these results in Table~\ref{tab:simulations_results} shows that, for \eq{}, a comparatively lower number of users submitted their credentials with respect to \eb{} (around \smamath{33\%} less for \cm{} and \cs{}). This result may suggest that even though quishing emails have the same effectiveness in terms of bringing a potential victim to a phishing webpage, such a victim may be somewhat more reluctant to submit their credentials. A possible explanation of this result, however, lies in the experimental setup of our experiments. Users were ultimately required to type their userid and passwords, which could be stored in a password manager accessible only from, e.g., the laptop or desktop used for work. If this is true, then less users submitted their credentials for \eq{} (w.r.t. \eb{}) because such users simply \textit{could not do so}---given that the landing webpage was visited on a smartphone, i.e., the device used to scan the QR code.

\textbf{Reported emails.} 
Let us provide some remarks on the number of ``reported'' emails of our simulation---especially with regard to \cs{}. Recall that, across \eb{}/\eq{}/\el{}, 11 employees of \cs{} reached out to the IT managers about the ``phishing'' emails they had just received (note: \smamath{60} emails were  sent in \cs{}). What is intriguing, however, is that \cs{} does not have a dedicated security team (see §\ref{ssec:companies}). Likely, such employees were somewhat suspicious of the email and opted for the easiest way of support they were aware of: contacting the most IT-savvy person in the company to ask for advice. Such an occurrence highlights the benefits of a low hierarchy organization (typical of small companies) and their close-knit structures of communication on phishing susceptibility. This aligns with the findings of Burda et al.~\cite{burda2023peculiar}, that phishing attacks towards SMEs can be stopped by the high level of direct communication, thus leading to users being alerted by coworkers quickly after an attack has been discovered. Comparatively speaking, \smamath{21.5\%} (resp. \smamath{40.2\%}) of the emails sent in \ch{} (resp. \cm) had been reported. Intriguingly, some prior works carried out in different contexts found that the ``report ratio'' of phishing emails (in a simulation) tends to be much lower---typically below 10\% (e.g.,~\cite{burda2020don,lain2024content}). A potential explanation lies in the heterogeneity of the reporting ecosystem across companies~\cite{sun2024victims}: for instance, a company that makes it easy to report emails (e.g., via a dedicated button) and that encourages their employees to be suspicious is likely to have a higher report rate~\cite{schiller2024employees}. Evidence that this is the case can be found in the results of our PPA questionnaire: many of our participants (across all companies) thought that a benign link was actually suspicious---which reflects an overly skeptical attitude. Hence, our numbers suggest that our considered companies promote reporting of suspicious emails.

\textbf{Some potential countermeasures.} Let us expand our suggestions in Section~§\ref{ssec:future} with additional insight and justifications.
\begin{itemize}[leftmargin=*]
    \item \textit{Defenses against quishing emails} are tough to realize. The issue is that implementation of automated mechanisms that can reliably detect the presence of (malicious) QR-codes in emails is a hard problem~\cite{ford2024feasibility}. This is because it is not known, a priori, if an email contains a QR-code---and, if so, where (e.g., it can be in an attachment, or embedded as an HTML object). Therefore, such mechanisms would necessitate a thorough scan of every email received by a given user, which would pose a lot of stress to the respective servers. Client-side solutions (e.g., implementing QR-code scanners with automated mechanisms that warn a user of a suspicious URL~\cite{rafsanjani2023qsecr}; or dedicated browser-extensions that perform their analyses at the client level) may not present substantial computational overhead, but may not be universally applicable, and/or may induce software lock-in. Nevertheless, as also recommended by~\cite{sharevski2022phishing}, we advocate future work to emphasize the importance of ``quishing education/training'': users should be aware that QR-codes may conceal cyber threats, and hence users should not blindly trust (and visit) the URLs interpreted by any given QR-code scanner.
    
    \item \textit{Dealing with malicious content generated by (OSINT-fed) LLMs} is an open issue. Humans can hardly distinguish human- from LLM-generated content~\cite{frank2024representative}. In the context of (OSINT-fed) LLM-written phishing emails, a plausible mitigation entails using detectors of machine-generated text (e.g.,~\cite{hans2024spotting}). For instance, if {\small \textit{(i)}}~a company tells its employees that company-related emails are not written by LLMs, then {\small \textit{(ii)}}~an automated mechanism that warns an employee that a given email contains LLM-written text which allegedly is company-related would induce the user to be more suspicious of the legitimacy of such an email. 
\end{itemize}
We stress that the aforementioned mitigations (which are based on our educated guesses) \textit{should not be taken as universal solutions to these problems}. Firstly, because they present tradeoffs; secondly, because they can be exploited by attackers (e.g., detectors of machine-generated text are not perfect, and require constant updates to be able to detect content generated by state-of-the-art LLMs).

\begin{table*}[hbtp!]    
    
    \centering
    \caption{\textbf{Questionnaire for measuring the PPA.} \textmd{Some questions (e.g., KCG1--5) have been provided with links or emails that were specific to the corresponding company (we cannot provide more details due to NDA). We did not use CSA1 in our main paper because, ultimately, nobody filled it for \cs[\scriptsize].}}
    \label{tab:questionnaire_1}
    \tiny
    \vspace{-4mm}
    \resizebox{1.99\columnwidth}{!}{
\begin{tabular}{{|>{\centering\arraybackslash}p{0.01\linewidth}|>{\centering\arraybackslash}p{0.13\linewidth}|>{\centering\arraybackslash}p{0.03\linewidth}|>{\centering\arraybackslash}p{0.28\linewidth}|>{\centering\arraybackslash}p{0.28\linewidth}|}} 
    \toprule
         \textbf{ID}&  \textbf{Category}&  \textbf{Item}&  \textbf{Question (English version)} &  \textbf{Question (German version)}\\ \midrule
         1
&  {\tiny Attitude towards Cybersecurity}
&  ACS1
&  I believe cybersecurity is important for protecting my personal information and online accounts.&  Ich glaube, dass Cybersicherheit wichtig ist, um meine persönlichen Daten und Online-Konten zu schützen.
\\ \hline 
         2
&  Attitude towards Cybersecurity
&  ACS2
&  I feel confident in my ability to identify cyber threats.&  Ich fühle mich sicher in meiner Fähigkeit, Cyber-Bedrohungen zu erkennen.
\\ \hline 
         3
&  Attitude towards Cybersecurity
&  ACS3
&  I feel confident in my ability to protect myself from cyber threats.&  Ich habe Vertrauen in meine Fähigkeit, mich vor Cyber-Bedrohungen zu schützen.
\\ \hline 
         4
&  Attitude towards Cybersecurity
&  ACS4
&  I believe that everyone has a role to play in protecting against cyber threats.&  Ich glaube, dass jeder eine Rolle beim Schutz vor Cyber-Bedrohungen spielen muss.
\\ \hline 
         5
&  Attitude towards Cybersecurity
&  ACS5
&  I feel a sense of responsibility to protect myself and others from cyber threats.&  Ich fühle mich dafür verantwortlich, mich und andere vor Cyber-Bedrohungen zu schützen.
\\ \hline 
         6
&  Attitude towards Cybersecurity
&  ACS6
&  I believe that staying informed about cybersecurity helps me to react effectively to unexpected situations.&  Ich glaube, dass es mir hilft, auf unerwartete Situationen effektiv zu reagieren, wenn ich über Cybersicherheit informiert bin.
\\ \hline 
         7
&  Attitude towards Cybersecurity
&  ACS7
&  I believe that cyber threats are becoming more common and sophisticated.&  Ich glaube, dass Cyber-Bedrohungen immer häufiger und raffinierter werden.
\\ \hline 
         8
&  Attitude towards Cybersecurity
&  ACS8
&  I am willing to take steps to improve my cybersecurity practices.&  Ich bin bereit, Maßnahmen zu ergreifen, um meine Cybersicherheitspraktiken zu verbessern.
\\ \hline 
         9
&  Attitude towards Cybersecurity
&  ACS9
&  I am willing to adapt my cybersecurity practices to new threats and challenges.&  Ich bin bereit, meine Cybersicherheitspraktiken an neue Bedrohungen und Herausforderungen anzupassen.
\\ \hline 
         10
&  Self-reported Behavior
&  BHV1
&  I believe that increased awareness of phishing scams would help to reduce the overall level of cybersecurity risky behavior.&  Ich glaube, dass eine stärkere Sensibilisierung für Phishing-Betrügereien dazu beitragen würde, das allgemeine Risikoverhalten im Bereich der Cybersicherheit zu verringern.
\\ \hline 
         11
&  Self-reported Behavior
&  BHV2
&  I believe that I am less likely to click on suspicious links or open attachments in emails because I am afraid of being phished.&  Ich glaube, dass ich weniger wahrscheinlich auf verdächtige Links klicke oder Anhänge in E-Mails öffne, weil ich Angst habe, Opfer eines Phishings zu werden.
\\ \hline 
         12
&  Self-reported Behavior
&  BHV3
&  My understanding of phishing scams affects my overall behavior when working with a digital device.&  Mein Verständnis von Phishing-Betrug beeinflusst mein allgemeines Verhalten bei der Arbeit mit einem digitalen Gerät.
\\ \hline 
         13
&  Self-reported Behavior
&  BHV4
&  I am confident in my ability to identify phishing emails.&  Ich habe Vertrauen in meine Fähigkeit, Phishing-E-Mails zu erkennen.
\\ \hline 
         14
&  Self-reported Behavior
&  BHV5
&  I recognized and avoided a phishing scams at least once in the past thanks to my phishing education.&  Ich habe in der Vergangenheit mindestens einmal einen Phishing-Betrug dank meiner Phishing-Aufklärung erkannt und vermieden.
\\ \hline 
         15
&  CSA Training experience
&  CSA1
&  Have you ever participated in an organizational cybersecurity training?&  Haben Sie jemals an einer organisatorischen Cybersicherheitsschulung teilgenommen?
\\ \hline 
         16
&  CSA Training experience
&  CSA2
&  I regularly complete my organization's cybersecurity awareness training.&  Ich nehme regelmäßig an den Cybersicherheitsschulungen meiner Organisation teil.
\\ \hline 
         17
&  Training Usability
&  TUB1
&  I believe that the information presented in my organization's cybersecurity awareness training was relevant and applicable to my work or personal life.&  Ich bin der Meinung, dass die Informationen, die in den Schulungen meines Unternehmens zum Thema Cybersicherheit vermittelt wurden, für meine Arbeit relevant und anwendbar waren.
\\ \hline 
         18
&  Training Usability
&  TUB2
&  I feel more knowledgeable about cybersecurity threats and prevention methods since participating in my organization's cybersecurity awareness training.&  Ich fühle mich besser informiert über Bedrohungen der Cybersicherheit und Präventionsmethoden, seit ich an der Schulung zum Thema Cybersicherheit in meiner Organisation teilgenommen habe.
\\ \hline 
         19
&  Training Usability
&  TUB3
&  I feel more confident in my ability to protect myself and my organization from cyber threats since participating in my organization's cybersecurity awareness training.&  Ich bin zuversichtlicher, dass ich mich und mein Unternehmen vor Cyber-Bedrohungen schützen kann, seit ich an der Schulung teilgenommen habe.
\\ \hline 
         20
&  Training Usability
&  TUB4
&  I am more likely to apply the knowledge and skills I learned in my organization's cybersecurity awareness training to my future cybersecurity practices.&  Ich werde das Wissen und die Fähigkeiten, die ich in der Schulung zum Bewusstsein für Cybersicherheit in meiner Organisation gelernt habe, in meinen zukünftigen Cybersecurity-Praktiken eher anwenden.
\\ \hline 
         21
&  Training Usability
&  TUB5
&  I feel more prepared to protect myself from online threats after learning about phishing in my company's cybersecurity awareness training.&  Ich fühle mich besser vorbereitet, mich vor Online-Bedrohungen zu schützen, nachdem ich in der Schulung meines Unternehmens über Phishing gelernt habe.
\\ \hline 
         22
&  Training Usability
&  TUB6
&  I believe that the skills I learned in my organization's cybersecurity awareness training will help me to better identify and respond to cybersecurity threats.&  Ich glaube, dass die Fähigkeiten, die ich in der Cybersecurity-Schulung meines Unternehmens gelernt habe, mir helfen werden, Cybersecurity-Bedrohungen besser zu erkennen und auf sie zu reagieren.
\\ \hline 
         23
&  Training Usability
&  TUB7
&  The topics covered in my organization's cybersecurity awareness training were relevant to my work and private life.&  Die Themen, die in der Schulung zum Thema Cybersicherheit in meinem Unternehmen behandelt wurden, waren für meine Arbeit und mein Privatleben relevant.
\\ \hline 
         24
&  Training Usability
&  TUB8
&  I found my organization's cybersecurity awareness training to be informative and engaging.&  Ich fand die Schulung zum Thema Cybersicherheit in meinem Unternehmen informativ und ansprechend.
\\ \hline 
         25
&  Knowledge and competence gain
&  KCG1
&  How suspicious are you of this link? (malicious link)&  Wie misstrauisch sind Sie gegenüber diesem Link?
\\ \hline 
         26
&  Knowledge and competence gain
&  KCG2
&  How suspicious are you of this link? (malicious link 2) &  Wie misstrauisch sind Sie gegenüber diesem Link?
\\ \hline 
         27
&  Knowledge and competence gain
&  KCG3
&  How suspicious are you of this link? (benign link) &  Wie misstrauisch sind Sie gegenüber diesem Link?
\\ \hline 
         28
&  Knowledge and competence gain
&  KCG4
&  How suspicious are you of this email? (benign email) &  Wie misstrauisch sind Sie gegenüber dieser E-Mail?
\\ \hline 
         29
&  Knowledge and competence gain
&  KCG5
&  How suspicious are you of this email? (malicious email) &  Wie misstrauisch sind Sie gegenüber dieser E-Mail?
\\ \hline 
         30
&  Knowledge and competence gain
&  KCG6
&  An email from my colleague cannot be a phishing email.&  Eine E-Mail von meinem Kollegen kann keine Phishing-E-Mail sein.
\\ \hline 
         31
&  Knowledge and competence gain
&  KCG7
&  Phishing emails always contain grammatical errors or poor spelling.&  Phishing-E-Mails enthalten immer grammatikalische Fehler oder schlechte Rechtschreibung.
\\ \hline 
         32
&  Knowledge and competence gain
&  KCG8
&  Phishing scams are not only a threat to people who use personal computers; mobile devices are susceptible to phishing attacks, too.&  Phishing-Betrügereien sind nicht nur eine Bedrohung für Menschen, die einen Computer benutzen; auch mobile Geräte sind anfällig für Phishing-Angriffe.
\\ \hline 
         33
&  Knowledge and competence gain
&  KCG9
&  Phishing scams are not only used to steal financial information; they are used to steal other types of data, such as personal information or login credentials.&  Phishing-Betrügereien dienen nicht nur dazu, finanzielle Informationen zu stehlen, sondern auch andere Arten von Daten, z. B. persönliche Informationen oder Anmeldedaten.
\\ \hline 
         34
&  Knowledge and competence gain
&  KCG10
&  Phishing can only occure if I am clicking on a link.&  Phishing kann nur stattfinden, wenn ich auf einen Link klicke.
\\ \hline 
         35
&  Socio-Demographics
&  SDG1
&  What is your age?&  Wie alt sind Sie?
\\ \hline 
         36
&  Socio-Demographics
&  SDG2
&  What is the highest degree or level of education you have completed?&  Welchen höchsten Abschluss haben Sie erreicht?
\\ \hline 
         37
&  Socio-Demographics
&  SDG3
&  How many years have you been with your current company?&  Wie viele Jahre arbeiten Sie bereits bei Ihrem derzeitigen Unternehmen?
\\ \hline 
         38
&  Socio-Demographics
&  SDG4
&  How many years of work experience do you have in total (not just at your current company)?&  Wie viele Jahre Berufserfahrung haben Sie insgesamt (nicht nur in Ihrem jetzigen Unternehmen)?
\\ \hline 
         39
&  Socio-Demographics
&  SDG5
&  How often do you use a digital device (e.g. Laptop, Desktop PC, Smartphone) for doing your work?&  Wie oft benutzen Sie ein digitales Gerät (z. B. Laptop, Desktop-PC, Smartphone) für Ihre Arbeit?
\\ \hline          
         40&  Socio-Demographics&  SDG6&  Which department do you work in?&  In welcher Abteilung arbeiten Sie?\\ \hline
    \end{tabular}
    }
    \end{table*}

\section{User Surveys and Questionnaires}
\label{app:questionnaires}

\noindent
Here, we provide additional information on our questionnaires.

\subsection{Implementation}
\label{sapp:questionnaire_implementation}

\begin{figure}[!htbp]
    \vspace{-3mm}
    \centering
    \includegraphics[width=0.9\columnwidth]{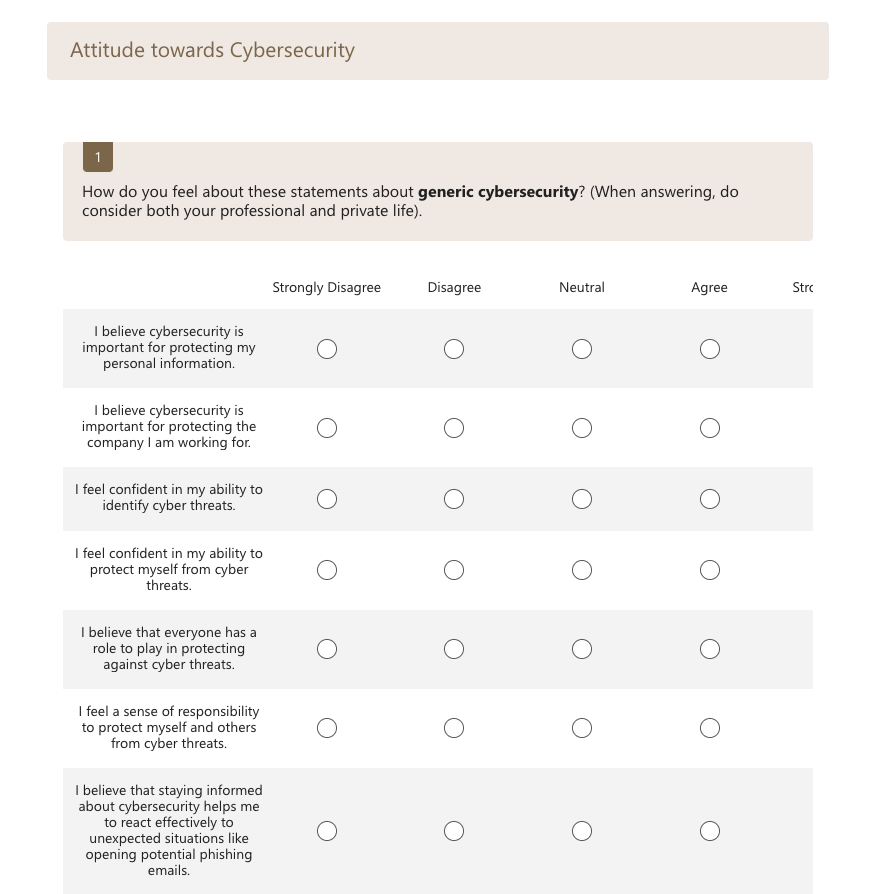}
    \vspace{-2mm}
    \caption{\textbf{Snippet of the ``Attitude towards Cybersecurity'' section of questionnaire.} \textmd{Every question could be answered in a 5-point Likert scale.}}
    \label{fig:survey_attitude}
    \vspace{-5mm}
\end{figure}

\begin{figure}[!htbp]
    \centering
    \includegraphics[width=0.85\columnwidth]{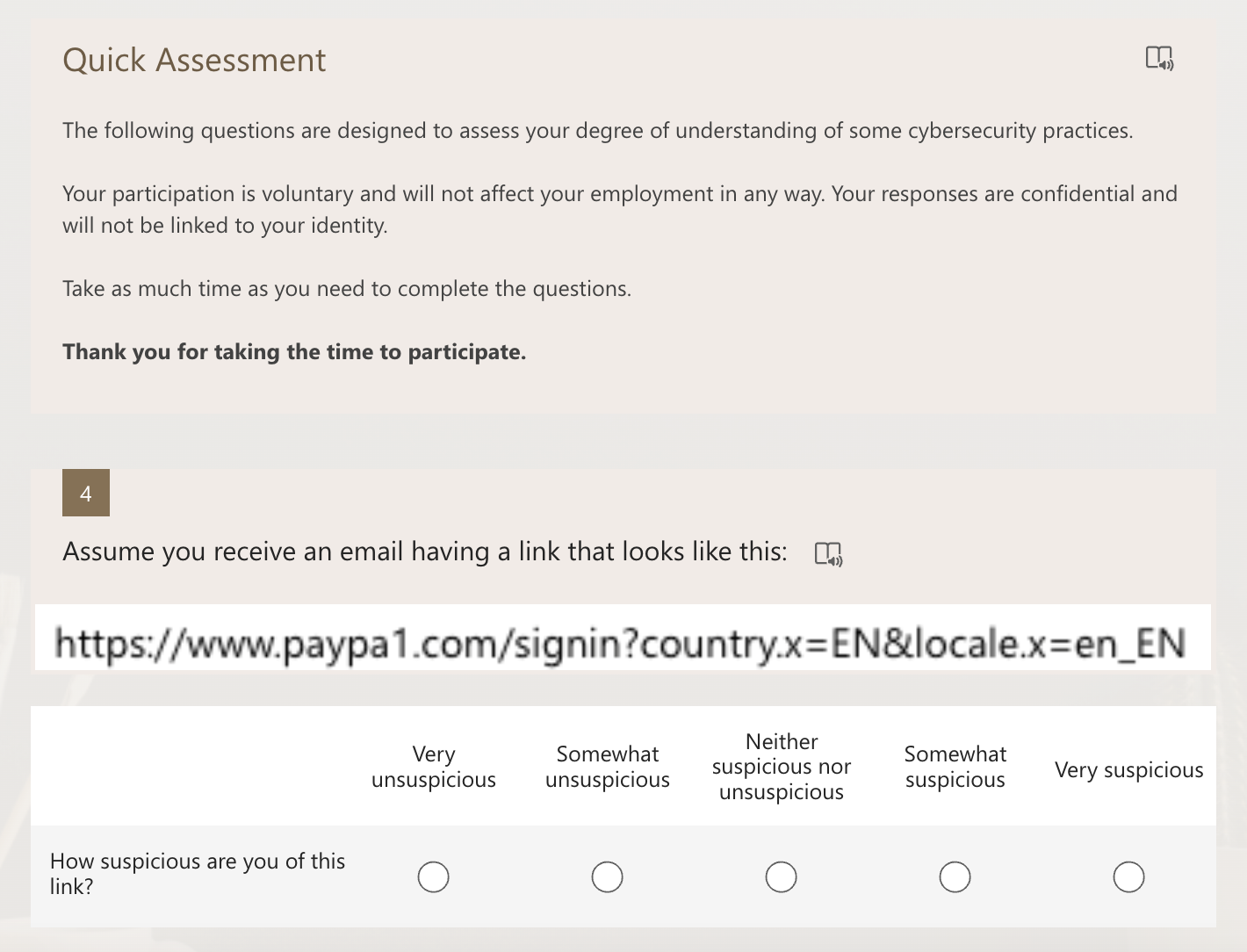}
    \vspace{-2mm}
    \caption{\textbf{Snippet of the ``quick assessment'' section of our questionnaire.} \textmd{Links could be different for every company. We also warn the user that an incorrect answer would not affect their employment.}}
    \label{fig:survey_quick}
    \vspace{-3mm}
\end{figure}

\noindent
We provide in Table~\ref{tab:questionnaire_1} (done with the employees) and Table~\ref{tab:questionnaire_2} (done with executives/managers of the companies) the complete questionnaires used in the user surveys with our companies. We also show snippets of questionnaire with employees in Fig.~\ref{fig:survey_attitude} and Fig.~\ref{fig:survey_quick}.

We stress that, before analysing our data to answer our third RQ, we performed \textit{preprocessing operations}. For instance, we manually checked if the employees could correctly identify phishing links, and hence had the required knowledge to answer the questions in the sections of ``competence and knowledge gain'' (KCG). Moreover, for items KCG3, KCG4, KCG8, and KCG9, we reversed the five-point Likert scale because these were benign examples, and respondents were expected to not be suspicious of these emails or statements. 

We did not include ``attention-check'' questions, as the questionnaire was intentionally short and participation was voluntary, reducing the likelihood of disengaged responses. However, we manually reviewed all submissions for inconsistencies and found none. On average, participants took approximately 10 minutes to complete the questionnaire, with no anomalous deviations. As a result, no responses were discarded.

\begin{table}[hbtp!]    
    \centering
    \caption{\textbf{Questionnaire with the companies' representatives.} \textmd{We used these answers to derive a profile of these companies (§\ref{ssec:companies}). Altogether, these questions also enable one to derive the remaining four indicators proposed by Chaudhary et al.~\cite{chaudhary2022developing} to investigate the ``impact'' factor (i.e., \textit{value added}, \textit{reachability}, \textit{touchability}, \textit{overall feedback}---see §\ref{ssec:awareness_implementation}).}}
    \label{tab:questionnaire_2}
    \vspace{-4mm}
    \resizebox{0.8\columnwidth}{!}{
\begin{tabular}{>{\centering\arraybackslash}p{0.1\linewidth}|>{\centering\arraybackslash}p{0.5\linewidth}|>{\centering\arraybackslash}p{0.2\linewidth}} 
    \toprule 
         \textbf{ID} &  \textbf{Question} & \textbf{Type} \\ 
    \midrule
         1&  Do you carry out cybersecurity awareness trainings?
& Single Choice\\ \hline 
         2&  How often do you train your emloyees?
& Short Answer\\ \hline 
         3&  When did you start training your employees?
& Short Answer\\ \hline 
 4& What are the topics which are being taught in the cybersecurity awareness trainings?
&Long Answer\\ \hline 
         5&  How often do you update the training?
& Short Answer\\ \hline 
         6&  Do you implement recent threats / attack trends into the training?
& Short Answer\\ \hline 
         7&  Which methods of delivery are being used for the training?
& Multiple Choice\\ \hline 
         8&  Do you differentiate the training's content among different target groups?
& Short Answer\\ \hline 
         9&  What kind of feedback do you receive for the cybersecurity awareness training?
& Single Choice\\ \hline 
 10& How many phishing simulations do you carry out each year?
&Short Answer\\ \hline 
         11&  Do you implement recent threats / attack trends into the simulations?& Short Answer\\ \bottomrule
    \end{tabular}
    }

\end{table}

\subsection{Demographics (PPA questionnaire)}
\label{sapp:demographics}

\noindent
We report in Tables~\ref{tab:demo_age}--\ref{tab:demo_workexp} the complete demographic details of the participants of our PPA-related questionnaire~(§\ref{ssec:awareness_implementation}).

\begin{table}[!htbp]
    \centering
    \caption{\textbf{Demographics: Age.}}
    \vspace{-3mm}
    \resizebox{0.5\columnwidth}{!}{
        \begin{tabular}{c|c|c|c}
            \toprule
            \textbf{Age range} & \cs{} & \cm{} & \ch{}\\
            \midrule
            
            18--24 years & $0$ & $7$ & $3$ \\
            25--34 years & $4$ & $22$ & $21$ \\
            35--44 years & $4$ & $19$ & $7$ \\
            45--54 years & $4$ & $17$ & $5$ \\
            55+ years & $1$ & $16$ & $0$ \\
            not provided & $0$ & $1$ & $0$ \\

            \bottomrule
        \end{tabular}
    }
    \label{tab:demo_age}
    \vspace{-1mm}
\end{table}

\begin{table}[!htbp]
    \centering
    \caption{\textbf{Demographics: Highest level of education.}}
    \vspace{-3mm}
    \resizebox{0.5\columnwidth}{!}{
        \begin{tabular}{c|c|c|c}
            \toprule
            \textbf{Education} & \cs{} & \cm{} & \ch{}\\
            \midrule
            
            High School & $1$ & $11$ & $1$ \\
            Bachelor's Degree & $5$ & $14$ & $6$ \\
            Master's Degree & $2$ & $31$ & $29$ \\
            PhD or higher & $0$ & $3$ & $0$ \\
            Other & $4$ & $21$ & $0$ \\
            not provided & $1$ & $2$ & $0$ \\

            \bottomrule
        \end{tabular}
    }
    \label{tab:demo_education}
    \vspace{-1mm}
\end{table}

\begin{table}[!htbp]
    \centering
    \caption{\textbf{Demographics: Area of Work.}}
    \vspace{-3mm}
    \resizebox{0.5\columnwidth}{!}{
        \begin{tabular}{c|c|c|c}
            \toprule
            \textbf{Area} & \cs{} & \cm{} & \ch{}\\
            \midrule
            
            Administration \& Support & $6$ & $8$ & $3$ \\
            Finance, Risk \& Audit & $0$ & $14$ & $0$ \\
            Human Resources & $0$ & $3$ & $0$ \\
            IT \& Digital Banking & $0$ & $18$ & $18$ \\
            Legal \& Compliance & $0$ & $8$ & $0$ \\
            Logistics & $0$ & $0$ & $2$ \\
            Management & $1$ & $1$ & $3$ \\
            Marketing \& Communications & $0$ & $2$ & $6$ \\
            Operations & $4$ & $19$ & $1$ \\
            not provided & $2$ & $9$ & $3$ \\

            \bottomrule
        \end{tabular}
    }
    \label{tab:demo_field}
    \vspace{-1mm}
\end{table}

\begin{table}[!htbp]
    \centering
    \caption{\textbf{Demographics: Work-related usage of digital devices.}}
    \vspace{-3mm}
    \resizebox{0.5\columnwidth}{!}{
        \begin{tabular}{c|c|c|c}
            \toprule
            \textbf{percentage} & \cs{} & \cm{} & \ch{}\\
            \midrule
            
            0--25\% of the time & $2$ & $0$ & $0$ \\
            26--50\% of the time & $1$ & $2$ & $0$ \\
            51--75\% of the time & $2$ & $5$ & $2$ \\
            75+\% of the time & $8$ & $75$ & $34$ \\
            not provided & $0$ & $0$ & $0$ \\

            \bottomrule
        \end{tabular}
    }
    \label{tab:demo_digital}
    \vspace{-1mm}
\end{table}

\begin{table}[!htbp]
    \centering
    \caption{\textbf{Demographics: Years of affiliation to the same company.}}
    \vspace{-3mm}
    \resizebox{0.5\columnwidth}{!}{
        \begin{tabular}{c|c|c|c}
            \toprule
            \textbf{Affiliation} & \cs{} & \cm{} & \ch{}\\
            \midrule
            
            0--2 years & $2$ & $32$ & $17$ \\
            3--5 years & $4$ & $16$ & $10$ \\
            6--10 years & $4$ & $10$ & $4$ \\
            10+ years & $2$ & $23$ & $5$ \\
            not provided & $1$ & $1$ & $0$ \\

            \bottomrule
        \end{tabular}
    }
    \label{tab:demo_affiliation}
    \vspace{-1mm}
\end{table}

\begin{table}[!htbp]
    \centering
    \caption{\textbf{Demographics: Work Experience in Years.}}
    \vspace{-3mm}
    \resizebox{0.5\columnwidth}{!}{
        \begin{tabular}{c|c|c|c}
            \toprule
            \textbf{Experience} & \cs{} & \cm{} & \ch{}\\
            \midrule
            
            0--2 years & $0$ & $2$ & $8$ \\
            3--5 years & $0$ & $8$ & $4$ \\
            6--10 years & $0$ & $8$ & $4$ \\
            10+ years & $12$ & $64$ & $11$ \\
            not provided & $1$ & $0$ & $0$ \\

            \bottomrule
        \end{tabular}
    }
    \label{tab:demo_workexp}
    \vspace{-1mm}
\end{table}

\subsection{Detailed Results (PPA questionnaire)}
\label{sapp:results}

\begin{figure}[!t]
    \centering
    \includegraphics[width=0.5\columnwidth]{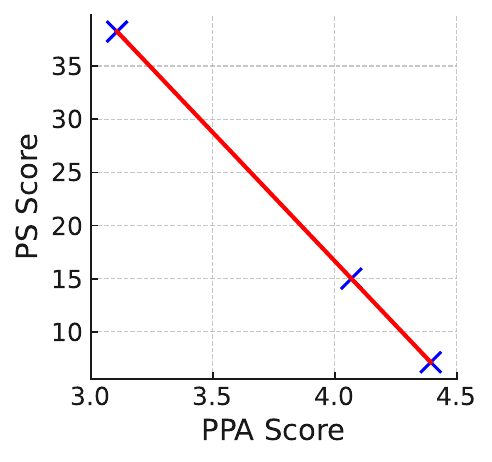}
    \vspace{-4mm}
    \caption{\textbf{Regression model of the perceived phishing awareness (PPA) w.r.t. phishing susceptibility (PS) for each company.} \textmd{The blue markers denote our datapoints, whereas the red line is the fitted regression model.}}
    \label{fig:linear}
    \vspace{-2mm}
\end{figure}

\noindent
We provide in Tables~\ref{tab:ppa_attitude}--\ref{tab:ppa_knowledge} the aggregated results of every question asked in our PPA-related questionnaire (refer to Table~\ref{tab:questionnaire_1} for the mapping between ``ItemCode'' and actual question). We also report in Fig.~\ref{fig:linear} the regression line of our statistical test (see §\ref{ssec:rq3}).

\begin{table}[!htbp]
    \centering
    \caption{\textbf{PPA assessment: Attitude towards Cybersecurity.}}
    \vspace{-3mm}
    \resizebox{0.99\columnwidth}{!}{
        \begin{tabular}{c?c|c||c|c||c|c}
            \toprule
            & \multicolumn{2}{c||}{\cs{}} & \multicolumn{2}{c||}{\cm{}} & \multicolumn{2}{c}{\ch{}}\\
            \cline{2-7}
            ItemCode & Mean & Var. & Mean & Var. & Mean & Var. \\
            \midrule
            
            ACS1 & $4.692$ & $0.213$ & $4.768$ & $0.178$ & $4.917$ & $0.076$\\
            ACS2 & $4.769$ & $0.178$ & $4.890$ & $0.098$ & $5.000$ & $0.000$\\
            ACS3 & $3.615$ & $1.006$ & $3.716$ & $0.598$ & $4.111$ & $0.710$\\
            ACS4 & $3.154$ & $1.207$ & $3.457$ & $0.816$ & $3.972$ & $0.860$\\
            ACS5 & $4.385$ & $0.544$ & $4.695$ & $0.285$ & $4.694$ & $0.268$\\
            ACS6 & $4.385$ & $0.544$ & $4.333$ & $0.543$ & $4.611$ & $0.293$\\
            ACS7 & $4.000$ & $0.769$ & $4.476$ & $0.493$ & $4.750$ & $0.188$\\
            ACS8 & $4.692$ & $0.367$ & $4.805$ & $0.181$ & $4.861$ & $0.231$\\
            ACS9 & $4.154$ & $0.746$ & $4.366$ & $0.598$ & $4.667$ & $0.333$\\
            ACS10 & $4.385$ & $0.391$ & $4.476$ & $0.371$ & $4.722$ & $0.201$\\

            \rowcolor{gray!30} overall & $4.223$ & $0.596$ & $4.398$ & $0.416$ & $4.631$ & $0.316$\\

            \bottomrule
        \end{tabular}
    }
    \label{tab:ppa_attitude}
    
\end{table}

\begin{table}[!htbp]
    \centering
    \caption{\textbf{PPA assessment: Self-reported Behavior.}}
    \vspace{-3mm}
    \resizebox{0.99\columnwidth}{!}{
        \begin{tabular}{c?c|c||c|c||c|c}
            \toprule
            & \multicolumn{2}{c||}{\cs{}} & \multicolumn{2}{c||}{\cm{}} & \multicolumn{2}{c}{\ch{}}\\
            \cline{2-7}
            ItemCode & Mean & Var. & Mean & Var. & Mean & Var. \\
            \midrule
            
            BHV1 & $4.462$ & $0.556$ & $4.305$ & $0.456$ & $4.667$ & $0.278$\\
            BHV2 & $3.846$ & $1.207$ & $3.963$ & $0.889$ & $4.528$ & $0.305$\\
            BHV3 & $3.538$ & $0.710$ & $3.732$ & $0.635$ & $4.111$ & $0.710$\\
            BHV4 & $4.154$ & $0.438$ & $4.671$ & $0.294$ & $4.722$ & $0.256$\\
           
            \rowcolor{gray!30} overall & $4.000$ & $0.728$ & $4.168$ & $0.569$ & $4.507$ & $0.387$\\

            \bottomrule
        \end{tabular}
    }
    \label{tab:ppa_behavior}
    
\end{table}

\begin{table}[!htbp]
    \centering
    \caption{\textbf{PPA assessment: CSA Training Experience and Training Usability.} \textmd{Only one valid question refers to CSA training experience in our questionnaire, i.e., CSA2; the other codes here refer to Training Usability.}}
    \vspace{-3mm}
    \resizebox{0.99\columnwidth}{!}{
        \begin{tabular}{c?c|c||c|c||c|c}
            \toprule
            & \multicolumn{2}{c||}{\cs{}} & \multicolumn{2}{c||}{\cm{}} & \multicolumn{2}{c}{\ch{}}\\
            \cline{2-7}
            ItemCode & Mean & Var. & Mean & Var. & Mean & Var. \\
            \midrule
            
            \rowcolor{gray!30} CSA2 & $1.667$ & $0.222$ & $4.091$ & $0.550$ & $4.667$ & $0.333$\\ \hline
            TUB1 & $3.333$ & $2.889$ & $4.260$ & $0.608$ & $4.583$ & $0.354$\\
            TUB2 & $2.333$ & $0.889$ & $3.779$ & $1.029$ & $4.306$ & $0.768$\\
            TUB3 & $2.667$ & $1.556$ & $3.792$ & $0.892$ & $4.194$ & $0.768$\\
            TUB4 & $3.000$ & $2.667$ & $4.273$ & $0.536$ & $4.306$ & $0.712$\\
            TUB5 & $3.000$ & $2.667$ & $3.816$ & $0.940$ & $4.278$ & $0.756$\\ 
            TUB6 & $3.000$ & $2.667$ & $3.948$ & $0.777$ & $4.278$ & $0.766$\\ 
            TUB7 & $2.667$ & $1.556$ & $4.117$ & $0.675$ & $4.444$ & $0.469$\\
            TUB8 & $2.333$ & $0.889$ & $3.870$ & $0.918$ & $4.417$ & $0.521$\\

            \rowcolor{gray!30} overall & $2.792$ & $1.972$ & $3.982$ & $0.797$ & $4.351$ & $0.638$\\

            \bottomrule
        \end{tabular}
    }
    \label{tab:ppa_training}
    
\end{table}

\begin{table}[!htbp]
    \centering
    \caption{\textbf{PPA assessment: Knowledge and Competence Gain.} \textmd{The codes KCG1--KCG5 refer to ``email and link identification'', whereas KCG6--KCG10 refer to ``knowledge assessment''.}}
    \vspace{-3mm}
    \resizebox{0.99\columnwidth}{!}{
        \begin{tabular}{c?c|c||c|c||c|c}
            \toprule
            & \multicolumn{2}{c||}{\cs{}} & \multicolumn{2}{c||}{\cm{}} & \multicolumn{2}{c}{\ch{}}\\
            \cline{2-7}
            ItemCode & Mean & Var. & Mean & Var. & Mean & Var. \\
            \midrule
            
            KCG1 & $4.769$ & $0.178$ & $4.768$ & $0.398$ & $4.778$ & $0.340$\\
            KCG2 & $3.692$ & $0.828$ & $3.646$ & $1.351$ & $4.361$ & $0.564$\\
            KCG3 & $2.154$ & $1.207$ & $2.580$ & $1.651$ & $2.944$ & $1.886$\\
            KCG4 & $2.462$ & $2.710$ & $2.695$ & $1.700$ & $3.000$ & $2.111$\\
            KCG5 & $4.769$ & $0.178$ & $4.573$ & $0.757$ & $4.444$ & $1.247$\\
            KCG6 & $1.923$ & $1.302$ & $3.976$ & $1.268$ & $4.417$ & $0.465$\\
            KCG7 & $2.462$ & $1.325$ & $3.988$ & $0.866$ & $4.056$ & $1.386$\\
            KCG8 & $1.462$ & $1.172$ & $4.354$ & $1.326$ & $4.500$ & $1.250$\\
            KCG9 & $1.417$ & $1.243$ & $4.463$ & $1.468$ & $4.417$ & $1.576$\\
            KCG10 & $3.154$ & $2.438$ & $3.793$ & $1.189$ & $4.111$ & $0.932$\\

            \rowcolor{gray!30} overall & $2.862$ & $1.258$ & $3.884$ & $1.197$ & $4.103$ & $1.176$\\

            \bottomrule
        \end{tabular}
    }
    \label{tab:ppa_knowledge}
    
\end{table}

\end{document}